\documentclass[10pt,5p]{elsarticle}

    \usepackage{jhoagg}
    
    \usepackage{mathtools}
    \usepackage{amsfonts}
    \usepackage{amssymb,latexsym}
    \usepackage[psamsfonts]{eucal}
    \usepackage{amsthm}
    \usepackage{graphicx}
    \usepackage{cleveref}

    \usepackage[inline]{enumitem}
    \usepackage{indentfirst}
    \usepackage{setspace}
    \usepackage{microtype}
    \usepackage{threeparttable}
    \usepackage{float}
    \usepackage{cleveref}
    \usepackage{afterpage}
    \usepackage{placeins}
    \usepackage{cancel}
    \usepackage{leftidx}
    \usepackage{lipsum}
    \usepackage{multicol}
    \usepackage{breqn}
    \usepackage[export]{adjustbox}
    \usepackage{comment}
    \usepackage[ruled, linesnumbered, lined]{algorithm2e}
    \usepackage[dvipsnames]{xcolor}
    \usepackage{xcolor}
    \usepackage[most]{tcolorbox}
    \usepackage{soul}
    \usepackage{standalone}
    \usepackage{textgreek}
    \usepackage{multirow}
    \usepackage{anyfontsize}
    \usepackage{tabularx}
    \newcolumntype{Y}{>{\centering\arraybackslash}X}
\usepackage{subcaption}

    \usepackage{cleveref}
    \usepackage{mathrsfs}

    \usepackage{etoolbox}
\makeatletter
\patchcmd{\@makecaption}
  {\scshape}
  {}
  {}
  {}
\makeatother

    \usepackage{tikz}
    \usetikzlibrary{shapes,arrows,positioning,fit}

    \usepackage{pgf}
    \usepackage{pgfplots}
    \usepgflibrary{plothandlers}
    \usepgfplotslibrary{external} 
    \pgfkeys{/pgf/number format/.cd,1000 sep={}}  
    \usetikzlibrary{calc,shapes.misc,plotmarks}
    \pgfplotsset{compat=1.13}
    
    \let\originalleft\left
    \let\originalright\right
    \renewcommand{\left}{\mathopen{}\mathclose\bgroup\originalleft}
    \renewcommand{\right}{\aftergroup\egroup\originalright}

    \newcounter{thm} 
    \theoremstyle{definition}
    \newtheorem{theorem}[thm]{\indent Theorem}
    
    \newtheorem{assumption}{\indent Assumption}
    
    \theoremstyle{definition}
    \newtheorem{proposition}{\indent Proposition}    
    
    \newtheorem{lemma}{\indent Lemma}
    
    \newtheorem{remark}{\indent Remark}
    
    \newtheorem{corollary}{\indent Corollary}
    
    \newtheorem{definition}{\indent Definition}
    
    \newtheorem{example}{\indent Example}
    
    \newtheorem{fact}{\indent Fact}
    
    \newtheorem{conjecture}{\indent Conjecture}
    
    \newtheorem{experiment}{ Experiment}

\newcommand{\nbproof}[1] {\vspace{5pt}\indent{\bf Proof of {#1}. }}
\newcommand{\neproof}{\vspace{5pt}}
    \allowdisplaybreaks

    \newlist{enumA}{enumerate}{1}
    \setlist[enumA,1]{label=(A\arabic*),leftmargin=1cm}

    \newlist{enumC}{enumerate}{1}
    \setlist[enumC,1]{label=(C\arabic*),leftmargin=1cm}

    		\newcommand\xqed[1]{%
      \leavevmode\unskip\penalty9999 \hbox{}\nobreak\hfill
      \quad\hbox{#1}}
    \newcommand\exampletriangle{\xqed{$\triangle$}}

    \newcommand\proofsquare{\xqed{$\square$}}

    \usepackage{accents}
    
    \newlength\figureheight 
    \newlength\figurewidth

    \DeclareMathOperator{\sgn}{sgn}

    \allowdisplaybreaks
    
    \graphicspath{ {Figures/} }

    \DeclareMathAlphabet{\mathcal}{OMS}{cmsy}{m}{n} 

    \crefname{equation}{}{}
    
    \usepackage[ruled, linesnumbered, lined]{algorithm2e}
    \SetKwInput{KwInit}{Initialize}
    \SetKwFunction{SafeSet}{SafeSet}
    \SetKwFunction{GetSafe}{GetSafe}
    \SetKwFunction{GetUnsafe}{GetUnsafe}
    
    \SetCommentSty{mycommfont}
    
    \newlist{enumalph}{enumerate}{1}
    \setlist[enumalph]{label=\textit{(\alph*)}}
    \usepackage{url}
    \usepackage{geometry}
\makeatletter
\def\ps@pprintTitle{%
     \let\@oddhead\@empty
     \let\@evenhead\@empty
     \let\@evenfoot\@empty
     \def\@oddfoot{\reset@font\small\parbox[b]{\textwidth}{\vspace{-2ex}© 2026. This manuscript version is made available under the 
    CC-BY-NC-ND 4.0 license \url{https://creativecommons.org/licenses/by-nc-nd/4.0/}}\hfil}%
}
\makeatother

\begin{document}

\begin{frontmatter}
\title{\LARGE Experimental Demonstration of a Decentralized Electromagnetic Formation Flying Control Using Alternating Magnetic Field Forces}
\author{Sumit S. Kamat, Ajin Sunny, T. Michael Seigler, and Jesse B. Hoagg\corref{cor1}}

\address{Department of Mechanical and Aerospace Engineering, University of Kentucky, Lexington, KY 40506-0503}

 \cortext[cor1]{Corresponding author; email: jesse.hoagg@uky.edu}

\begin{abstract}
Electromagnetic formation flying (EMFF) is challenging due to the complex coupling between the electromagnetic fields generated by each satellite in the formation. 
To address this challenge, this article uses alternating magnetic field forces (AMFF) to decouple the electromagnetic forces between each pair of satellites.
The key idea of AMFF is that a pair of alternating (e.g., sinusoidal) magnetic moments results in a nonzero time-averaged interaction force if and only if those alternating magnetic moments have the same frequency.
Hence, the approach in this article is to drive each satellite's electromagnetic actuation system with a sum of sinusoids, where each frequency is common to only a pair of satellites. 
Then, the amplitudes of each sinusoid are modulated (i.e., controlled) to achieve the desired forces between each pair of satellites.
The main contribution of this article is an experimental demonstration of 3-satellite decentralized closed-loop EMFF using AMFF.
To the authors' knowledge, this is the first demonstration of AMFF with at least 3 satellites in open or closed loop. 
This is noteworthy because the coupling challenges of EMFF are only present with more than 2 satellites, and thus, a formation of at least 3 is necessary to evaluate the effectiveness of AMFF.
The experiments are conducted on a ground-based testbed consisting of 3 electromagnetically actuated satellites on linear air tracks. 
{The closed-loop experiments demonstrate decentralized EMFF with AMFF where the maximum steady-state formation error is less than $\pm 0.01$~m and the settling time is less than $30$~s.
These experiments validate the decoupling of intersatellite forces through frequency-multiplexed AMFF.}
The closed-loop experimental results are compared with the behavior of numerical simulations. 
\end{abstract}

\end{frontmatter}

{
\section{Notation} \label{sec:notation}

Physical vectors are denoted with bold symbols, for example, $\mathbf{r}$. 
The magnitude of $\mathbf{r}$ is denoted by $| \mathbf{r} |$.
A frame is a collection of mutually-orthogonal physical unit vectors.

Let $\mathbb{N}$ denote the set of nonnegative integers.
Let $\mathcal I \triangleq \{1, ..., n\}$, where $n \in \mathbb{N}$ is the number of satellites in the formation, and $\mathcal P \triangleq \{(i,j) \in \mathcal I \times \mathcal I : i \neq j\}$, which is the set of ordered pairs.
Unless otherwise stated, all statements in this paper that involve the subscript $i$, $ij$, and $k$ are for all $i \in \mathcal I$, all $(i,j) \in \mathcal{P}$, and all $k \in \mathbb{N}$.
}

{
\section{Nomenclature}

\begin{tabbing}
XXXXXXXXX \= \kill 

$\mathcal{F} = \begin{bmatrix}
    \mathbf{i} &\mathbf{j} &\mathbf{j}
\end{bmatrix}$ \> Inertial frame\\

$\mathbf{r}_{i}$ \> Position of satellite $i$ (m)\\

$\mathbf{v}_i$ \> Velocity of satellite $i$ (m/s)\\

$\mathbf{r}_{ij} \triangleq \mathbf{r}_i - \mathbf{r}_j$ \> {Position of satellite $i$ relative to $j$} (m)\\

$\mathbf{F}_{ij}$ \> Force applied to satellite $i$ by $j$ (N)\\

$\mathbf{f}$ \> Intersatellite force function (A$^2$-m$^4$)\\

$\mu_{0}$ \> Vacuum permeability constant (H/m)\\

$m$ \> Mass of each satellite (kg)\\

$\mathbf{u}_i$ \> Magnetic moment of satellite $i$ (A-m$^2$)\\

$\mathbf{d}_{ij}$ \> Desired relative position (m)\\

$\omega_{ij}$ \> Interaction frequency (rad/s)\\

$\mathbf{p}_{ij,k}$ \> Amplitude control (A-m$^2$)\\

$T$ \> Update period for control (s)\\

$\bar{\mathbf{F}}_{ij}$ \> Time-averaged intersatellite force (N)\\

$\hat{\mathbf{F}}_{ij}$ \> Approximate time-averaged \\ \> intersatellite force (N)\\

$\mathbf{r}_{ij,k}$ \> Sampled relative position (m)\\

$\mathbf{v}_{ij,k}$ \> Sampled relative velocity (m/s)\\

$\mathbf{f}^*_{ij,k}$ \> Desired value of $\mathbf{f}(\mathbf{r}_{ij,k},\mathbf{p}_{ij,k},\mathbf{p}_{ji,k})$ (A$^2$-m$^4$)\\

$\alpha_{ij}$, $\rho_{ij}$, $\beta$ \> Control gains\\ 

$r_{ij,k}$ \> Relative position in $\mathbf{i}$ direction (m)\\

$v_{ij,k}$ \> Relative velocity in $\mathbf{i}$ direction (m/s)\\

$d_{ij}$ \> Desired relative position in $\mathbf{i}$ direction (m)\\

$f^*_{ij,k}$ \> $\mathbf{f}^*_{ij,k}$ in $\mathbf{i}$ direction (A$^2$-m$^4$)\\

$p_{ij,k}$ \> Amplitude control in $\mathbf{i}$ direction (A-m$^2$)\\

$I_{i}$ \> Current in the coil of satellite $i$ (A)\\

$I_{ij,k}$ \> Current amplitude (A)\\

$f$ \> $\mathbf{f}$ in $\mathbf{i}$ direction  (A$^2$-m$^4$)\\

$\xi_{ij,k}$ \> Integrator state (m)\\

$\bar{I}$ \> Maximum allowable current (A)\\

$\hat{F}(r_{ij},p_{ij}, p_{ji})$ \> $\hat{\mathbf{F}}_{ij}$ in $\mathbf{i}$ direction (N)\\

$q_{ij,k}$ \> Measurement of $r_{ij}$ (m)\\

$\hat{r}_{ij,k}$ \> Estimate of $r_{ij}$ (m)\\

$\hat{v}_{ij,k}$ \> Estimate of $v_{ij}$ (m/s)\\

$V$ \> Variance of noise associated with $q_{ij,k}$ (m$^2$)\\

$L$ \> Kalman filter gain\\

$P$ \> Solution to algebraic Riccati equation\\

$w$ \>  Variance of disturbance force (m$^2$/s$^4$)\\

$\hat{r}_{\mathrm{os}}$ \>  Overshoot (m)\\

$T_{\rms}$ \>Settling time (s)\\

$P_{\hat F}$ \>   Root mean square of $\hat{F}$ (N)
\end{tabbing}}

\section{Introduction} \label{sec:Introduction}

Spacecraft formation flying can advance a variety of space technologies such as distributed-aperture telescopes, gravity-wave detectors, and interferometers.
Multiple spacecraft working cooperatively are often capable of meeting or exceeding the capabilities of one monolithic spacecraft. 
Multi-satellite missions generally require formation flying, where satellites coordinate their position \cite{Beard2001,Lee2015,Ren2004,Ren2007,Scharf2003,Scharf2004} and potentially their attitude \cite{Lawton2002, Dimaragonas2009, Sarlette2009, SarletteSIAM2009, Ren2010, Du2016, Wang1999, Chavan2023}.
One challenge for spacecraft formation flying is that traditional propellant thrusters eventually deplete, and they can contaminate sensitive spacecraft components.

Electromagnetic formation flying (EMFF) is accomplished using electromagnetic coils that are onboard satellites in a formation. 
Each satellite's coils generate a magnetic field, which interacts with the magnetic fields of the other satellites to create magnetic field forces.
These magnetic field forces can control the relative positions of satellites \cite{Kong2004,KWON2010,Porter2014}. 
An advantage of EMFF is that the power source is renewable, while conventional actuation systems such as propellant thrusters eventually deplete.

EMFF for a single pair of satellites where the coils are driven by direct current (DC) is addressed in \cite{Elias2007,Kwon2011}. 
Experiments using DC-driven coils are presented in \cite{Kwon2011}; this includes open-loop experiments, as well as position-hold and trajectory follow-up experiments. 
In addition, feasibility experiments onboard the International Space Station are presented in \cite{Porter2014}, where attractive and repulsive forces were generated between 2 DC-driven coils.

EMFF for more than 2 satellites is challenging because the intersatellite forces are nonlinear functions of the magnetic moments generated by all satellites as well as the relative positions of all satellites. 
In other words, there is complex coupling between the electromagnetic fields generated by all satellites, and this coupling ultimately leads to intersatellite forces between every pair of satellites.
EMFF control for more than 2 satellites is considered in \cite{Ahsun2006,Schweighart2010,Cai2023}.
However, these approaches require centralization of all measurement information and solving nonconvex constrained optimization problems that do not scale well to a large number of satellites. 
An alternative approach in \cite{Riberos2010Th} relies on scheduling the actuation of each EMFF satellite over time. 
In this case, only a subset of the satellites are actuated for each time interval, which limits computational complexity because control computations are computed for only a subset of satellites. 
However, this approach requires centralization of the schedule and the control is computed through a nonconvex optimization. 
In sum, the coupling between electromagnetic fields causes scalability and decentralization challenges for EMFF.

A different approach to EMFF is to use alternating magnetic field forces (AMFF) to address the challenge of coupling between the electromagnetic fields.
The key idea of AMFF is that a pair of alternating (e.g., sinusoidal) magnetic moments results in a nonzero average interaction force between a pair of satellites if and only if those alternating magnetic moments have the same frequency~\cite{Youngquist2013,Nurge2016}.

A decentralized EMFF method using AMFF is presented in \cite{Abbasi2022}. 
In this method, each satellite computes its desired intersatellite forces using measurements of its position and velocity relative to only its local neighbor satellites, avoiding the centralization and scalability limitations discussed above.
Then, each satellite uses a sum of frequency-multiplexed sinusoidal magnetic moments to achieve desired intersatellite forces between each pair of satellites. 
For example, consider 3 satellites, where satellite~1 generates a magnetic moment that is the sum of sinusoids at 10~Hz and 20~Hz, satellite~2 generates a magnetic moment at 10~Hz, and satellite~3 generates a magnetic moment at 20~Hz. 
In this scenario, the average force between satellites~1 and 2 is a function of the amplitudes of the sinusoids at the common frequency 10~Hz.
Notably, the average force between satellites~1 and~2 is independent of the sinusoidal amplitudes at 20~Hz.
Similarly, the average force between satellites~1 and~3 is a function of the amplitudes at the common frequency 20~Hz.
Finally, since the magnetic moments of satellites~2 and~3 do not have a common frequency, the average force between these satellites is negligible. 
Thus, frequency-multiplexed AMFF uses a sum of sinusoidal magnetic moments on each satellite to decouple the time-averaged intersatellite forces.
Only pairs with a common frequency can generate a nonzero time-averaged intersatellite force, and that force is determined solely by the amplitudes at the common frequency.

In addition to \cite{Abbasi2022}, EMFF using AMFF has been explored in \cite{Huang2015,Zhang2016,Abbasi2020Scitech,Song2023,Song2023CCC,Takahasi2022}. 
For example, \cite{Huang2015,Zhang2016} presents a leader-following method with AMFF using sliding-mode control.
Relative position and attitude control with AMFF is addressed with a centralized algorithm in \cite{Takahasi2022} and using a decentralized algorithm in \cite{Abbasi2020Scitech}. 
In \cite{Song2023,Song2023CCC}, AMFF is combined with traditional reaction wheels in a centralized algorithm for relative position and attitude control in the presence of disturbance torques caused by interaction with Earth's magnetic field.

Experimental evaluations of EMFF with AMFF are in \cite{Nurge2016, Abbasi2020, Takahashi2025}.
Open-loop experiments demonstrating the AMFF approach are presented in \cite{Nurge2016}. 
Closed-loop single-degree-of-freedom experiments with a pair of satellites on linear air-tracks are presented in \cite{Abbasi2020,Takahashi2025}. 
Note that since AMFF uses sinusoidal currents, it can be implemented with standard coils (as demonstrated in \cite{Nurge2016, Abbasi2020, Takahashi2025}) as opposed to superconducting coils, which have been used in DC-actuation experiments.
Since the coupling challenges of EMFF are only present with more than 2 satellites, a formation of at least 3 satellites is necessary to evaluate the effectiveness of AMFF.
However, to the authors' knowledge, the existing literature does not include any experiments using AMFF with more than 2 satellites.

The main contribution of this article is an experimental demonstration of closed-loop EMFF using AMFF with 3 satellites.
This is the first demonstration of AMFF with at least 3 satellites in open or closed loop.
This work demonstrates EMFF with 3 satellites on a linear air track, where the coil of the center satellite is driven by a sum of 2 sinusoidal currents and each outer satellite is driven by one of the sinusoidal currents. 
As a result, intersatellite forces are generated between the center satellite and each outer satellite. 
This article adopts the piecewise-sinusoidal magnetic moment approach in \cite{Abbasi2022} to address the challenges of coupling between the electromagnetic fields; this approach is reviewed in \Cref{sec:Average_Satellite_Dynamics}. 
Then, \Cref{sec:Abbasi_Algorithm} presents the closed-loop decentralized EMFF algorithm, which is adopted from \cite{Abbasi2022} but with a simplified approach for the control amplitude allocation.
\Cref{sec:exp_setup} describes the experimental testbed and modifies the AMFF control to account for hardware limitations such as sensor noise and amplitude saturation in the control current. 
Experimental results are presented in \Cref{sec:Exp_Results_discussion}. 
First, this article presents open- and closed-loop experiments with 2 satellites to demonstrate the feasibility of EMFF with AMFF.
Next, this article presents closed-loop experiments with 3 satellites, where the satellites achieve formation in less than 30~s with maximum steady-state formation error less than $\pm 0.01$~m and the mean steady-state formation error less than $\pm 0.001$~m.
These one-dimensional experiments validate the decoupling of intersatellite forces through frequency-multiplexed AMFF.
This method extends to 3-dimensional position control by implementing at least 3 coils (orthogonal configuration); however, no additional actuation frequencies are needed and computation scales linearly with the number of satellites.  
The extension to 3-dimensional position-and-attitude control can be accomplished with one additional actuation frequency per interacting satellite pair (e.g., \cite{Abbasi2020Scitech}) and a doubling of computation.

\section{Method Overview}

The core concept of AMFF is that a pair of sinusoidal magnetic moments results in a nonzero time-averaged interaction force between a pair of satellites if and only if those alternating magnetic moments have the same frequency. 
The method in this paper is decentralized, which means that each satellite has access to measurements of its position and velocity relative to only its local neighbor satellites. 
First, this article presents a time-averaged model for the dynamics of a  satellite formation, where the controllable magnetic moments are a sum of amplitude-modulated sinusoids. 
Next, the article shows how the sinusoidal amplitudes are selected to achieve desired intersatellite forces, which are determined decentrally by each satellite using feedback of its neighbors.

The main contribution of this article is the first experimental demonstration of EMFF using AMFF with more than 2 satellites. 
First, open- and closed-loop experiments are conducted with 2 satellites to demonstrate feasibility of AMFF. 
Next, closed loop experiments with 3 satellites are performed to demonstrate the core concept of AMFF.


\begin{figure}[hbt]
    \centering
    \includegraphics[width=0.42\textwidth]{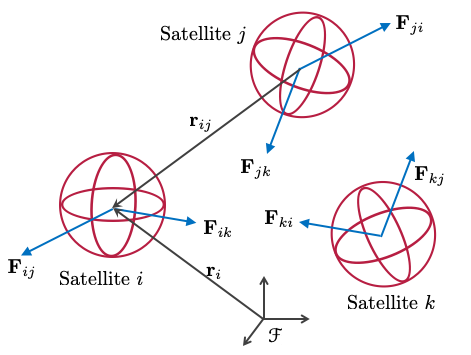}
    \caption{Each satellite is equipped with an electromagnetic actuation system consisting of three orthogonal coils. 
    The relative positions of the satellites are controlled by the interaction of equal-and-opposite intersatellite forces produced by the actuation systems.}
    \vspace{-10pt}
    \label{fig:Satellites_formation}
\end{figure}


\section{Satellite Dynamics} \label{sec:Dynamics}

Consider a system of $n$ satellites, where each satellite has mass $m$. 
The position $\mathbf r_i$ locates the mass center of satellite $i$ relative to the origin of an inertial frame $\CMcal{F}$, which consists of the right-handed set of mutually orthogonal unit vectors $\begin{bmatrix} \mathbf{i} &\mathbf{j} &\mathbf{k} \end{bmatrix}$. 
The velocity $\mathbf{v}_i$ and acceleration $\dot{\mathbf{v}}_i$ are the first and the second time-derivatives of $\mathbf{r}_i$ with respect to $\CMcal{F}$. The relative position $\mathbf{r}_{ij} \triangleq \mathbf{r}_i - \mathbf{r}_j$ locates the mass center of satellite $i$ relative to the mass center of satellite $j$. 
The relative velocity $\mathbf{v}_{ij}$ is the time derivative of $\mathbf{r}_{ij}$ with respect to $\CMcal{F}$. 
The satellite system is shown in \Cref{fig:Satellites_formation}.

Each satellite has an electromagnetic actuation system (i.e., multiple electromagnetic coils) that creates a magnetic field. 
These magnetic fields interact to produce intersatellite forces, which control the satellites' relative positions. 
The coils are modeled as magnetic dipoles, and the resulting force applied to satellite $i$ by satellite $j$ is
\begin{equation}
  \mathbf{F}_{ij} = \frac{3 \mu_0}{4 \pi |\mathbf r_{ij}|^4} \mathbf{f} (\mathbf{r}_{ij}, \mathbf{u}_i, \mathbf{u}_j), \label{F_ij}
\end{equation}
where $\mu_0$ is the vacuum permeability constant, $\mathbf{u}_i$ is the magnetic moment of satellite $i$, and the intersatellite force function is
\begin{align}
    \mathbf{f}(\mathbf{r}_{ij},&\mathbf{u}_{i},\mathbf{u}_j) \triangleq 
    \left( \mathbf{u}_j \cdot \frac{\mathbf{r}_{ij}}{|\mathbf{r}_{ij}|} \right) \mathbf{u}_i + 
 \left(\mathbf{u}_i \cdot \frac{\mathbf{r}_{ij}}{|\mathbf{r}_{ij}|} \right) \mathbf{u}_j 
 \notag
 \\
 &+ \left[(\mathbf{u}_i \cdot \mathbf{u}_j)-5 \left(\mathbf{u}_i \cdot \frac{\mathbf{r}_{ij}}{|\mathbf{r}_{ij}|} \right) \left(\mathbf{u}_j \cdot \frac{\mathbf{r}_{ij}}{|\mathbf{r}_{ij}|} \right) \right]\frac{\mathbf{r}_{ij}}{|\mathbf{r}_{ij}|}. \label{f(r_ui_uj)}
\end{align}
The magnetic moment $\mathbf{u}_i$, which is the control of satellite $i$, is a function of the current supplied to the electromagnetic coils. 
Since $\mathbf{r}_{ij} = -\mathbf{r}_{ji}$, it follows from \cref{F_ij,f(r_ui_uj)} that $\mathbf{F}_{ij} = - \mathbf{F}_{ji}$, that is, the force applied to satellite $j$ by $i$ is equal and opposite the force applied to $i$ by $j$. 
See \cite{Ahsun2006} and \cite{Schweighart2010} for more information on \cref{F_ij,f(r_ui_uj)} .

The translational dynamics of satellite $i$ are
\begin{equation}
    \dot{\mathbf{v}}_i \triangleq \frac{c_0}{m} \sum_{j \in \CMcal{I} \setminus \{ i \}}\frac{1}{|\mathbf{r}_{ij}|^4}\mathbf{f}(\mathbf{r}_{ij},\mathbf{u}_{i},\mathbf{u}_j),
    \label{accel_i}
\end{equation}
where $c_0 \triangleq 3 \mu_0 / (4 \pi)$. Since $\mathbf{F}_{ij} = - \mathbf{F}_{ji}$, it follows from \eqref{accel_i} that $\sum_{i \in \mathcal{I}} \dot{\mathbf{v}}_i (t)= 0$, which implies that the linear momentum of the system of satellites is conserved. 
Thus, the intersatellite electromagnetic forces can be used to alter relative positions, but they have no effect on the overall mass center of the satellites. 
Therefore, traditional actuation systems may be needed in combination with EMFF to control the formation's mass center.
For example, propellant thrusters may be needed for orbit maintenance in the presence of disturbances (e.g., atmospheric drag in low Earth orbit) or to allow for orbit transfers.
Similarly, a traditional attitude control system, such as reaction wheels, could be required for attitude control and addressing disturbances such as torques introduced by interactions with the Earth's magnetic field.

Let the constant $\mathbf{d}_{ij}$ be the desired position of satellite $i$ relative to satellite $j$, where $\mathbf{d}_{ij} = - \mathbf{d}_{ji}$. 
The objective is to design and demonstrate a decentralized control method that relies on feedback of neighboring satellites and yields $\lim_{t \to \infty} \mathbf{r}_{ij}(t) = \mathbf{d}_{ij}$.

\section{Time-Averaged Formation Dynamics with Sinusoidal Controls}
\label{sec:Average_Satellite_Dynamics}

This section reviews the piecewise-sinusoidal magnetic-moment approach in~\cite{Abbasi2022}, which is used to address the coupling that occurs between electromagnetic fields. 
This approach enables decentralized formation control, where each satellite uses relative position and velocity measurements of only its neighbors.
In this approach, each satellite uses a piecewise-sinusoidal magnetic moment $\mathbf{u}_i$ that is the sum of up to $n-1$ sinusoids with unique frequencies. 
Each unique frequency is common to only one pair of satellites. 
Thus, there are a total of up to $n(n-1)/2$ unique frequencies, and the amplitudes of each common-frequency pair of sinusoids are selected to prescribe the time-averaged intersatellite force between the associated satellite pair.

Let $\omega_{ij} \geq 0$ be the interaction frequency between satellite $i$ and $j$, where $\omega_{ij} = \omega_{ji}$ is unique if it is nonzero. 
If an interaction frequency is zero, then the time-averaged intersatellite force between the associated satellite pair is zero. 
Next, let $T>0$ be such that for all $(i,j) \in \mathcal{P}$, $\omega_{ij}T/2\pi$ is an integer. 
In other words, $T$ is an integer multiple of the periods of each sinusoid.
For all $k \in \mathbb{N}$ and $t \in [ kT, kT + T )$, consider the piecewise-sinusoidal control
\begin{equation}
    \mathbf{u}_i(t)=\sum_{j \in \mathcal{I} \setminus \{i\} } \mathbf p_{ij,k} \sin \omega_{ij}t
    \label{u_i}
\end{equation}
where the amplitudes $\{ \mathbf{p}_{ij,k} \}^{\infty}_{k=0}$ are the  control variables.

\begin{figure}[hbt]
    \centering
    \includegraphics[width=0.49\textwidth]{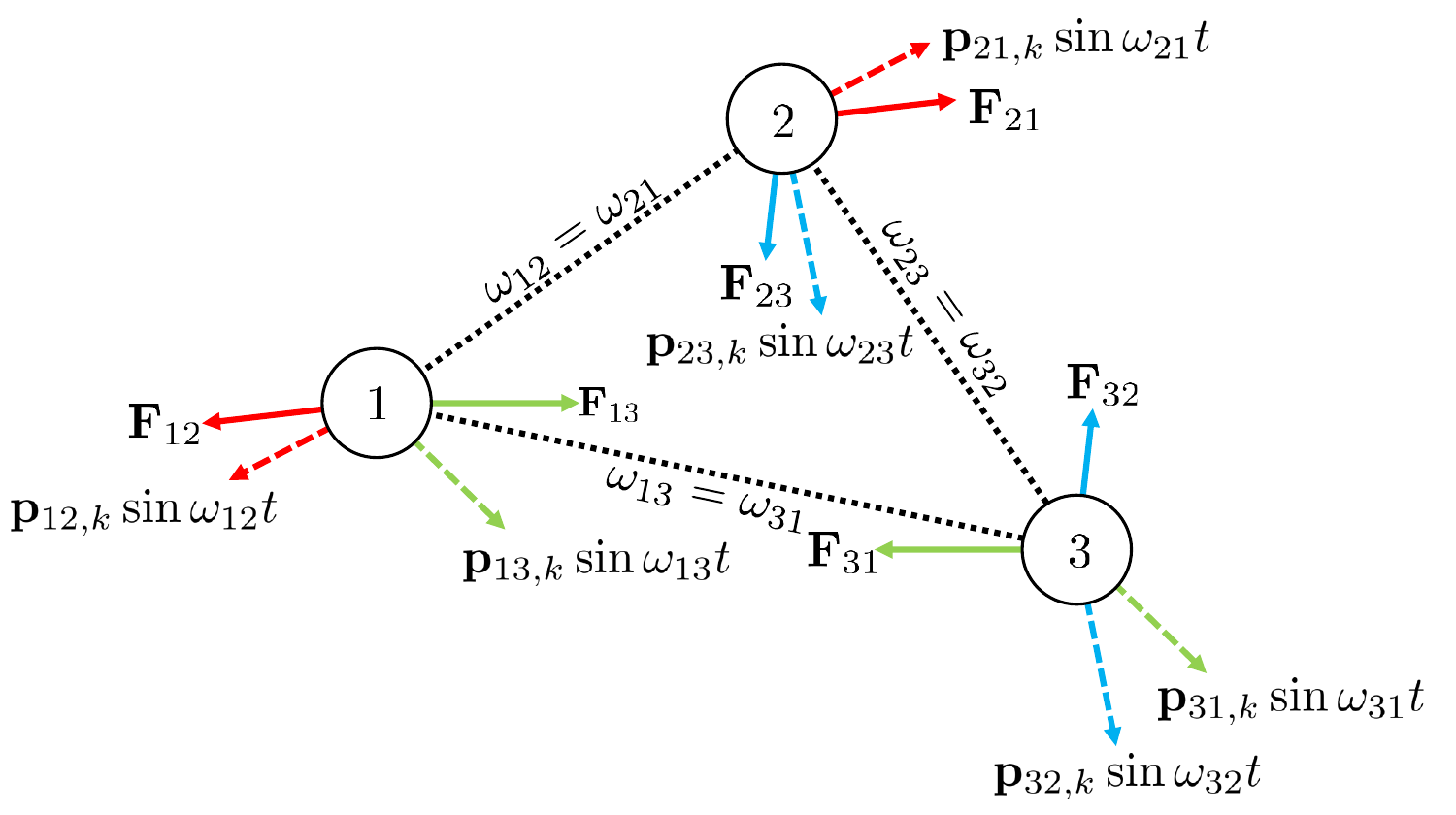}
    \caption{AMFF with 3 satellites and unique frequencies $\omega_{12}=\omega_{21}$, $\omega_{13}=\omega_{31}$, and $\omega_{23}=\omega_{32}$. 
    For each pair of satellites, the intersatellite forces ${\mathbf{F}}_{ij}=-{\mathbf{F}}_{ji}$ are determined on average from the amplitudes $(\mathbf{p}_{ij,k},\mathbf{p}_{ji,k})$ at frequency $\omega_{ij}$.
    For example, consider the case discussed in \Cref{sec:Introduction}, where $\omega_{12}=\omega_{21}=20\pi$~rad/s, $\omega_{13}=\omega_{31}=40\pi$~rad/s, and $\omega_{23}=\omega_{32}=0$~rad/s. 
    Then, the time-averaged force between satellites~1 and~2 is determined from amplitudes $(\mathbf{p}_{12,k},\mathbf{p}_{21,k})$, and the time-averaged force between satellites~1 and~3 is determined from amplitudes $(\mathbf{p}_{13,k},\mathbf{p}_{31,k})$. 
    In contrast, the time-averaged force between satellites~2 and~3 is zero because they do not have a common nonzero interaction frequency. 
}
    \label{fig:Moment_forces}
\end{figure}

The key idea of AMFF is that the amplitude pair $(\mathbf{p}_{ij,k},\mathbf{p}_{ji,k})$ at the common frequency $\omega_{ij}$ determines the time-averaged forces between satellites $i$ and $j$.
\Cref{fig:Moment_forces} illustrates this concept for 3 satellites, where the components of $\mathbf{u}_i$ and $\mathbf{u}_j$ at $\omega_{ij}$ cause the time-averaged forces between satellites $i$ and $j$.
The following result characterizes this decoupling property, which allows us to formalize the core concept of AMFF.
The result is from \cite[Proposition~1]{Abbasi2022}, and it is a consequence of averaging the product of sinusoids over an integer number of periods.

\begin{proposition} \label{Proposition_1}
Let $i, j \in \mathbb{N}$, and let $\mathbf{r}$ be a constant. Then, for all $k \in \mathbb{N}$,
\begin{equation*}
        \frac{1}{T} \int^{kT+T}_{kT} \mathbf f(\mathbf{r}, \mathbf{u}_i(t),\mathbf{u}_j(t)) \ \text{d}t \triangleq \frac{1}{2} \mathbf f(\mathbf{r}, \mathbf{p}_{ij,k},\mathbf{p}_{ji,k}).
\end{equation*}
\end{proposition}

We use \Cref{Proposition_1} to derive a time-averaged model of
the dynamics (\ref{accel_i}). 
Integrating (\ref{accel_i}) over the interval $[kT, kT + T ]$ yields
\begin{equation}
    \mathbf{v}_i(kT+T) =  \mathbf{v}_i(kT) + \frac{T}{m} \sum_{j \in \mathcal{I} \backslash \{ i \}} \bar{\mathbf{F}}_{ij} (k),
\label{discrete_velocity}    
\end{equation}
where the time-averaged intersatellite force is
\begin{equation*}
    \bar{\mathbf{F}}_{ij} (k) \triangleq \frac{1}{T} \int_{kT}^{kT+T} \frac{c_0}{| \mathbf{r}_{ij} (t) |^4} \mathbf{f}(  \mathbf{r}_{ij} (t), \mathbf{u}_{i} (t), \mathbf{u}_{j} (t)) \ \text{d}t.
\end{equation*}
We define the \textit{approximate time-averaged intersatellite force}
\begin{equation}
    \hat{\mathbf{F}}_{ij,k}  \triangleq \frac{1}{T} \int_{kT}^{kT+T} \frac{c_0}{| \mathbf{r}_{ij,k}  |^4} \mathbf{f}(  \mathbf{r}_{ij,k}, \mathbf{u}_{i} (t), \mathbf{u}_{j} (t)) \ \rmd t,
    \label{avg_approx_F_ij}
\end{equation}
which is an approximation of $\bar{\mathbf{F}}_{ij,k}$ obtained by replacing $\mathbf{r}_{ij}(t)$ with its sampling $\mathbf{r}_{ij,k} \triangleq \mathbf{r}_{ij}(kT)$. 
\Cref{Proposition_1} implies that 
\begin{equation}
    \hat{\mathbf{F}}_{ij,k} = \frac{c_0}{2 |\mathbf{r}_{ij,k}|^4 } \mathbf{f} (  \mathbf{r}_{ij,k}, \mathbf{p}_{ij,k} , \mathbf{p}_{ji,k}).
    \label{avg_approx_Fij_w_pij}
\end{equation}
Note that $(\hat{\mathbf{F}}_{ij,k}, \hat{\mathbf{F}}_{ji,k})$ depends on the amplitude control
pair $(\mathbf{p}_{ij,k} , \mathbf{p}_{ji,k})$, which does not affect the approximate time-averaged intersatellite force between any other pair of satellites. 
Thus, if $\mathbf{r}_{ij}$ does not change significantly over $[kT,kT+T)$, then the time-averaged formation dynamics can be approximated by \eqref{discrete_velocity} with 
$\bar{\mathbf{F}}_{ij} (k)$ replaced by $\hat{\mathbf{F}}_{ij,k}$.


\section{EMFF Control Algorithm}
\label{sec:Abbasi_Algorithm}

\subsection{Desired Intersatellite Forces}

The intersatellite feedback structure is described using an undirected graph. The vertex set of the undirected graph is $\mathcal{I}$. Let $\mathcal{E} \subset \mathcal{I} \times \mathcal{I}$ be the edge set. 
We assume that the undirected graph $\mathcal{G} = (\mathcal{I}, \mathcal{E})$ is connected. The neighbor set of satellite $i$ is $\mathcal{N}_i \triangleq \{ j \in \mathcal{I} : (i, j ) \in \mathcal{E} \}$, and satellite $i$ has access to $\{ \mathbf{r}_{ij,k} \}_{j \in \mathcal{N}_i}$ and $\{ \mathbf{v}_{ij,k} \}_{j\in \mathcal{N}_i}$ for feedback, where $\mathbf{v}_{ij,k} \triangleq \mathbf{v}_i(kT)-\mathbf{v}_j(kT)$.

Define the desired value for the intersatellite force function
\begin{equation}
    \mathbf{f}^*_{ij,k} \triangleq -\frac{ 2 m |\mathbf{r}_{ij,k}|^4 }{c_0} 
    \alpha_{ij} (( \mathbf{r}_{ij,k}-\mathbf{d}_{ij} ) + \beta \mathbf{v}_{ij,k} )
    \label{c_ij}
\end{equation}
where $ \alpha_{ij} = \alpha_{ji} \geq 0$ and $\beta>0$ are feedback gains. If $(i, j) \notin \mathcal{E}$, then $\alpha_{ij} = 0$; otherwise, $\alpha_{ij} > 0$.

If $\mathbf{f}(\mathbf{r}_{ij,k}, \mathbf{p}_{ij,k}, \mathbf{p}_{ji,k}) = \mathbf{f}^*_{ij,k}$, then it follows from \Cref{avg_approx_Fij_w_pij} that 
\begin{equation}
    \hat{\mathbf{F}}_{ij,k} = -m \alpha_{ij} \left[(\mathbf{r}_{ij,k} - \mathbf{d}_{ij}) + \beta \mathbf{v}_{ij,k} \right]. \label{eq:app_avg_force}
\end{equation}
In this case, \cite[Theorem 5]{Abbasi2022} provides conditions on selecting the gains $\alpha_{ij}$ and $\beta$ such that the formation error $\mathbf{r}_{ij,k}-\mathbf{d}_{ij}$ tends to zero for $\bar{\mathbf{F}}_{ij}(k)=\hat{\mathbf{F}}_{ij,k}$.
The approximate time-averaged intersatellite force \Cref{eq:app_avg_force} is based on linear consensus for sampled-data double integrators \cite{Cao2010}. This method uses feedback to determine approximate time-averaged intersatellite forces $\hat{\mathbf{F}}_{ij,k}$ that create virtual linear springs and dashpots between satellite $i$ and $j$.
We note that larger $\alpha_{ij}$ tends to result in faster formation convergence but can cause oscillations (e.g., overshoot) in the response.
Larger $\beta$ increases the virtual intersatellite damping and tends to reduce oscillations.
We also note that the analysis in \cite[Theorem 5]{Abbasi2022} relies on the assumption that $\bar{\mathbf{F}}_{ij}(k)=\hat{\mathbf{F}}_{ij,k}$, which is a reasonable approximation if $\mathbf{r}_{ij}$ does not change significantly over $[kT,kT+T)$.


\subsection{Control Amplitude Pair To Achieve Prescribed Intersatellite Force}

This section provides a construction for the amplitude pair $(\mathbf p_{ij,k},\mathbf p_{ji,k})$ such that $\mathbf f( \mathbf r_{ij,k}, \mathbf p_{ij,k}, \mathbf p_{ji,k}) =\mathbf{f}^{*}_{ij,k}$. 
This amplitude pair is constructed from $\mathbf{r}_{ij,k}$ and $\mathbf{f}^*_{ij,k}$.
The construction is presented in \cite{Kamat2025ACC,Kamat2025b} for physical vectors resolved in an inertial frame. 
Here, we present the construction coordinate free.
The following functions of a relative position $\mathbf{r}$ and desired force function value $\mathbf f^*$ are needed to construct the control amplitudes.
 Specifically, we define
\begin{align}
    \mathbf{g}(\mathbf r, \mathbf f^*) &\triangleq \begin{cases}
        \frac{g_{\mathrm{r}}(\mathbf r, \mathbf f^*)}{|\mathbf{r}|} \mathbf r +  \frac{g_{\mathrm{rf}}(\mathbf r, \mathbf f^*)}{|\mathbf r||\mathbf{r} \times \mathbf f^*|} ((\mathbf{r} \times \mathbf f^*) \times \mathbf r), & \mathbf r \times \mathbf f^* \neq \mathbf 0 ,
        \\
        \frac{g_{\mathrm{r}}(\mathbf r, \mathbf f^*)}{|\mathbf{r}|} \mathbf r, & \mathbf r \times \mathbf f^* = \mathbf 0, 
    \end{cases}
    \label{eq:g(r_c)}
    \\
    \mathbf{h}(\mathbf r, \mathbf f^*) &\triangleq \begin{cases}
        \frac{h_{\mathrm{r}}(\mathbf r, \mathbf f^*)}{|\mathbf{r}|} \mathbf r +  \frac{h_{\mathrm{rf}}(\mathbf r, \mathbf f^*)}{|\mathbf r||\mathbf{r} \times \mathbf f^*|} ((\mathbf{r} \times \mathbf f^*) \times \mathbf r), & \mathbf r \times \mathbf f^* \neq \mathbf 0 ,
        \\
        \frac{h_{\mathrm{r}}(\mathbf r, \mathbf f^*)}{|\mathbf{r}|} \mathbf r, & \mathbf r \times \mathbf f^* = \mathbf 0,
    \end{cases}
    \label{eq:h(r_c)}
\end{align}
where
\begin{align}
    g_{\mathrm{r}}(\mathbf r, \mathbf f^*) &\triangleq -\frac{\sgn(\mathbf r \cdot \mathbf f^*)}{2} \left( \frac{|\mathbf r \cdot \mathbf f^*| + \Phi_1(\mathbf r, \mathbf f^*)}{|\mathbf r|} \right)^{\frac{1}{2}}, \label{eq:gr}
    \\
    g_{\mathrm{rf}}(\mathbf r, \mathbf f^*) &\triangleq \frac{1}{\sqrt{2}} \left( \frac{-|\mathbf r \cdot \mathbf f^*| + \Phi_2(\mathbf r, \mathbf f^*)}{|\mathbf r|} \right)^{\frac{1}{2}}, \label{eq:grc}
    \\
    h_{\mathrm{r}}(\mathbf r, \mathbf f^*) &\triangleq \frac{1}{2} \left( \frac{|\mathbf r \cdot \mathbf f^*| + \Phi_2(\mathbf r, \mathbf f^*)}{|\mathbf r|} \right)^{\frac{1}{2}}, \label{eq:hr}
    \\
    h_{\mathrm{rf}}(\mathbf r, \mathbf f^*) &\triangleq -\frac{\sgn(\mathbf r \cdot \mathbf f^*) }{\sqrt{2}} \left( \frac{-|\mathbf r \cdot \mathbf f^*| + \Phi_1(\mathbf r, \mathbf f^*)}{|\mathbf r|} \right)^{\frac{1}{2}}, \label{eq:hrc}
\end{align}
and
\begin{align}  
    \Phi_1(\mathbf r, \mathbf f^*) &\triangleq \sqrt{ | \mathbf r \times \mathbf f^* |^2 + |\mathbf r|^2 |\mathbf f^*|^2 }, \label{eq:Phi1}
    \\
    \Phi_2(\mathbf r, \mathbf f^*) &\triangleq (2 - \sgn(\mathbf r \cdot \mathbf f^*)^2 ) \Phi_1(\mathbf r, \mathbf f^*). \label{eq:Phi2}
\end{align}
The following result shows that \Cref{eq:g(r_c),eq:h(r_c)} are an amplitude pair such that $\mathbf f$ takes on the prescribed value $\mathbf f^*$.

\begin{proposition}    \label{Proposition_2}
    For all $\mathbf f^*$ and all $\mathbf r \neq \mathbf 0$, $$\mathbf f(\mathbf r, \mathbf g(\mathbf r,\mathbf f^*), \mathbf h(\mathbf r, \mathbf f^*)) = \mathbf f^*.$$
\end{proposition}

\Cref{Proposition_2} shows how to select the amplitudes $\mathbf{p}_{ij,k}$ and $\mathbf{p}_{ji,k}$ of the common frequency sinusoids in \Cref{u_i}. 
In other words, \Cref{Proposition_2} implies that we can use \Cref{eq:g(r_c),eq:h(r_c)} to select $(\mathbf{p}_{ij,k}, \mathbf{p}_{ji,k})$ such that $\mathbf f( \mathbf r_{ij,k}, \mathbf p_{ij,k}, \mathbf p_{ji,k}) =\mathbf{f}^{*}_{ij,k}$, which implies that the approximate time-averaged intersatellite force equals the desired value given by \Cref{eq:app_avg_force}. 
Specifically, we let the amplitudes be
\begin{equation}
    \mathbf{p}_{ij,k} =\begin{cases}
            \mathbf g(\mathbf r_{ij,k}, \mathbf{f}^*_{ij,k}), &i<j,\\
             \mathbf h(\mathbf r_{ij,k}, \mathbf{f}^*_{ij,k}), &i>j.\\
        \end{cases}
        \label{eqn:p_ij_allocation}
\end{equation}
 It follows from \Cref{Proposition_2} that $\mathbf{f}(\mathbf{r}_{ij,k}, \mathbf{p}_{ij,k}, \mathbf{p}_{ji,k})= \mathbf{f}^*_{ij,k}$. 
 In other words, \Cref{Proposition_2} confirms that the amplitude pair $(\mathbf{p}_{ij,k},\mathbf{p}_{ji,k})$ given by \eqref{eqn:p_ij_allocation} achieves the desired intersatellite force, which confirms the core decoupling approach of AMFF.

\nbproof{Proposition~\ref{Proposition_2}}
We consider 2 cases: (i) $\mathbf r \times \mathbf f_* \neq \mathbf 0$ and (ii) $\mathbf r \times \mathbf f_* = \mathbf 0$.

First, consider (i) $\mathbf r \times \mathbf f_* \neq \mathbf 0$. 
Substituting $|\mathbf r \times \mathbf f_*| = |\mathbf r|^2 \mathbf f_* - (\mathbf r \cdot \mathbf f_*) \mathbf r$, \eqref{eq:g(r_c)}, and \eqref{eq:h(r_c)} into \eqref{f(r_ui_uj)} yields
\begin{align}
    \mathbf f(\mathbf r, \mathbf g, \mathbf h) =& \left( - \frac{2 g_{\mathrm r} h_{\mathrm r}}{|\mathbf{r}|} + \frac{g_{\mathrm{rf}} h_{\mathrm{rf}}}{|\mathbf{r}|} -  \frac{(g_{\mathrm{r}} h_{\mathrm{rf}}  + g_{\mathrm{rf}} h_{\mathrm{r}})(\mathbf r \cdot \mathbf f_*)}{|\mathbf{r}| |\mathbf r \times \mathbf f_*|}  \right) \mathbf r \notag 
    \\ 
    & + \frac{(g_{\mathrm{r}} h_{\mathrm{rf}}  + g_{\mathrm{rf}} h_{\mathrm{r}})|\mathbf r|}{|\mathbf r \times \mathbf f_*|} \mathbf f_*. \label{eq:f(gr_grc_hr_hrc)_r_c}
\end{align}
where the arguments $\mathbf r$ and $\mathbf f_*$ are omitted for brevity. 
We consider 2 subcases: $\mathbf r \cdot \mathbf f_* =0$ and $\mathbf r \cdot \mathbf f_* \neq 0$. 
For $\mathbf r \cdot \mathbf f_*=0$, it follows from \eqref{eq:gr}--\eqref{eq:hrc} that $g_{\mathrm{r}} = h_{\mathrm{rf}}=0$ and $g_{\mathrm{rf}} h_{\mathrm{r}} = |\mathbf r \times \mathbf f_*|/ |\mathbf r|$, and substituting into \eqref{eq:f(gr_grc_hr_hrc)_r_c} yields $\mathbf f(\mathbf r, \mathbf g, \mathbf h) = \mathbf f_*$. Next, for $\mathbf r \cdot \mathbf f_* \neq 0$, it follows from \eqref{eq:Phi1} and \eqref{eq:Phi2} that $\Phi_2=\Phi_1$, and using \eqref{eq:gr}--\eqref{eq:hrc} yields
\begin{align*}
    g_{\mathrm{r}} h_{\mathrm{r}} &= - \frac{\sgn(\mathbf r \cdot \mathbf f_*)}{4 |\mathbf r|} \left( |\mathbf r \cdot \mathbf f_*| + \Phi_1 \right),
    \\
    g_{\mathrm{r}} h_{\mathrm{rf}} &= g_{\mathrm{rf}} h_{\mathrm{r}} = \frac{|\mathbf r \times \mathbf f_*|}{2 |\mathbf r|},  
    \\
     a_y b_y &= -\frac{\sgn(\mathbf r \cdot \mathbf f_*) }{2 |\mathbf r|} \left( -|\mathbf r \cdot \mathbf f_*| + \Phi_1 \right).
\end{align*}
Thus, substituting into \eqref{eq:f(gr_grc_hr_hrc)_r_c} yields $\mathbf f( \mathbf r, \mathbf g, \mathbf h) =  \mathbf f_*$, which confirms the result for (i) $\mathbf r \times \mathbf f_* \neq 0$.

Next, consider (ii) $\mathbf r \times \mathbf f_* = \mathbf 0$, and substituting \eqref{eq:g(r_c)}--\eqref{eq:hrc} into \eqref{f(r_ui_uj)} yields
\begin{align}
    \mathbf f(\mathbf r, \mathbf g, \mathbf h) &= -2 g_{\mathrm{r}} h_{\mathrm{r}} \frac{ \mathbf r}{|\mathbf r|} \notag
    \\
    &= \frac{\sgn(\mathbf r \cdot \mathbf f_*)}{2 |\mathbf r|^2}(|\mathbf r \cdot \mathbf f_*| +\Phi_1)^{\frac{1}{2}}(|\mathbf r \cdot \mathbf f_*| +\Phi_2)^{\frac{1}{2}} \mathbf r. \label{eq:f(r,g,h)_r_x_c_neq_0}
\end{align}
We consider 2 subcases: $\mathbf r \cdot \mathbf f_* =0$ and $\mathbf r \cdot \mathbf f_* \neq 0$. 
For $\mathbf r \cdot \mathbf f_*=0$, it follows from \eqref{eq:f(r,g,h)_r_x_c_neq_0} that $\mathbf f(\mathbf r, \mathbf g, \mathbf h) =\mathbf 0$. 
Since $\mathbf r \cdot \mathbf f_*=0$ and $\mathbf r \times \mathbf f_* = \mathbf 0$, it follows that $\mathbf f_*= \mathbf 0$, which implies that $\mathbf f(\mathbf r, \mathbf g, \mathbf h) = \mathbf f_*$. 
Next, for $\mathbf r \cdot \mathbf f_* \neq 0$, it follows from \eqref{eq:Phi1} and \eqref{eq:Phi2} that $\Phi_2=\Phi_1$. 
Since $\mathbf r \cdot \mathbf f_* \neq 0$ and $\mathbf r \times \mathbf f_* = \mathbf 0$, it follows that $\mathbf r = (\sgn(\mathbf r \cdot \mathbf f_*) |\mathbf r|/ |\mathbf f_*| ) \mathbf f_*$ and $|\mathbf r \cdot \mathbf f_*| +\Phi_1 = 2 |\mathbf r| |\mathbf f_*|$.
Thus, substituting into \eqref{eq:f(r,g,h)_r_x_c_neq_0} yields $\mathbf f( \mathbf r, \mathbf g, \mathbf h) =  \mathbf f_*$, which confirms the result for (ii) $ \mathbf r \times \mathbf f_* = \mathbf 0$.
\proofsquare
\neproof


\section{Experiment Setup}
\label{sec:exp_setup}

This section describes a one-dimensional (1D) experimental platform for testing the EMFF algorithm from \Cref{sec:Average_Satellite_Dynamics,sec:Abbasi_Algorithm}, specializes the algorithm to 1D, and presents modifications for implementation.

\begin{figure}[hbt]
    \centering
    \includegraphics[width=0.485\textwidth]{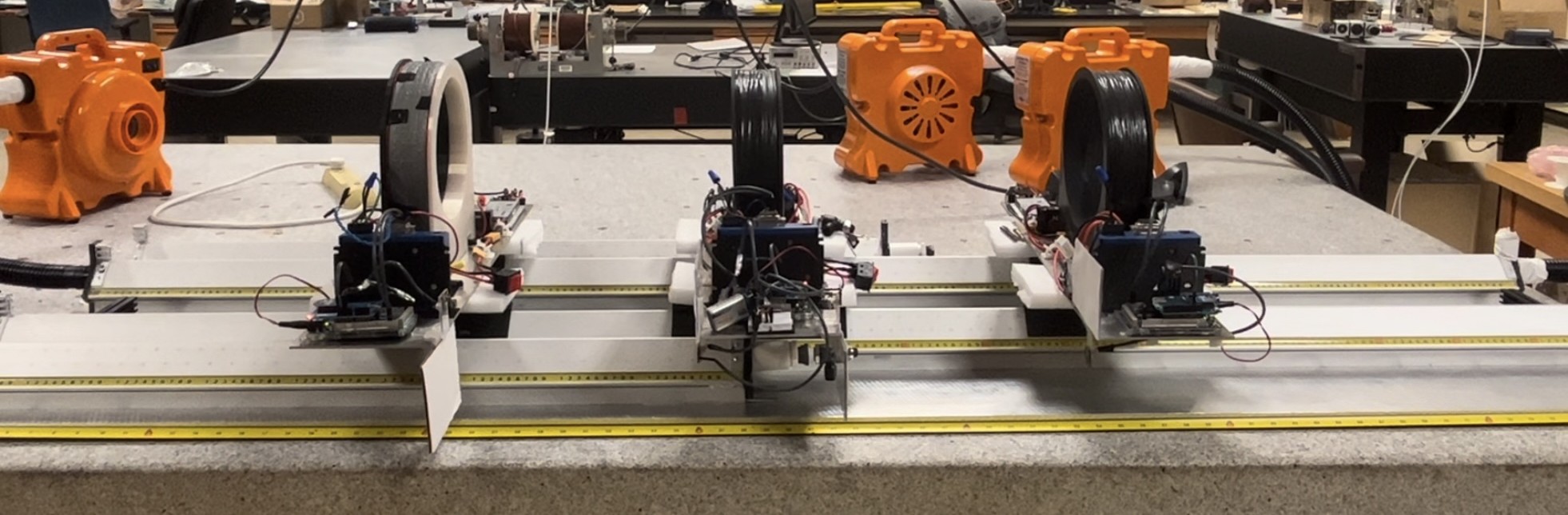}
    \caption{Experimental platform. Three EAS units sit on the linear air tracks, where the air is supplied by the air sources to create low-friction motion.}
    \label{fig:testbed}
\end{figure}

\begin{figure}[hbt]
    \centering
    \includegraphics[width=0.482\textwidth]{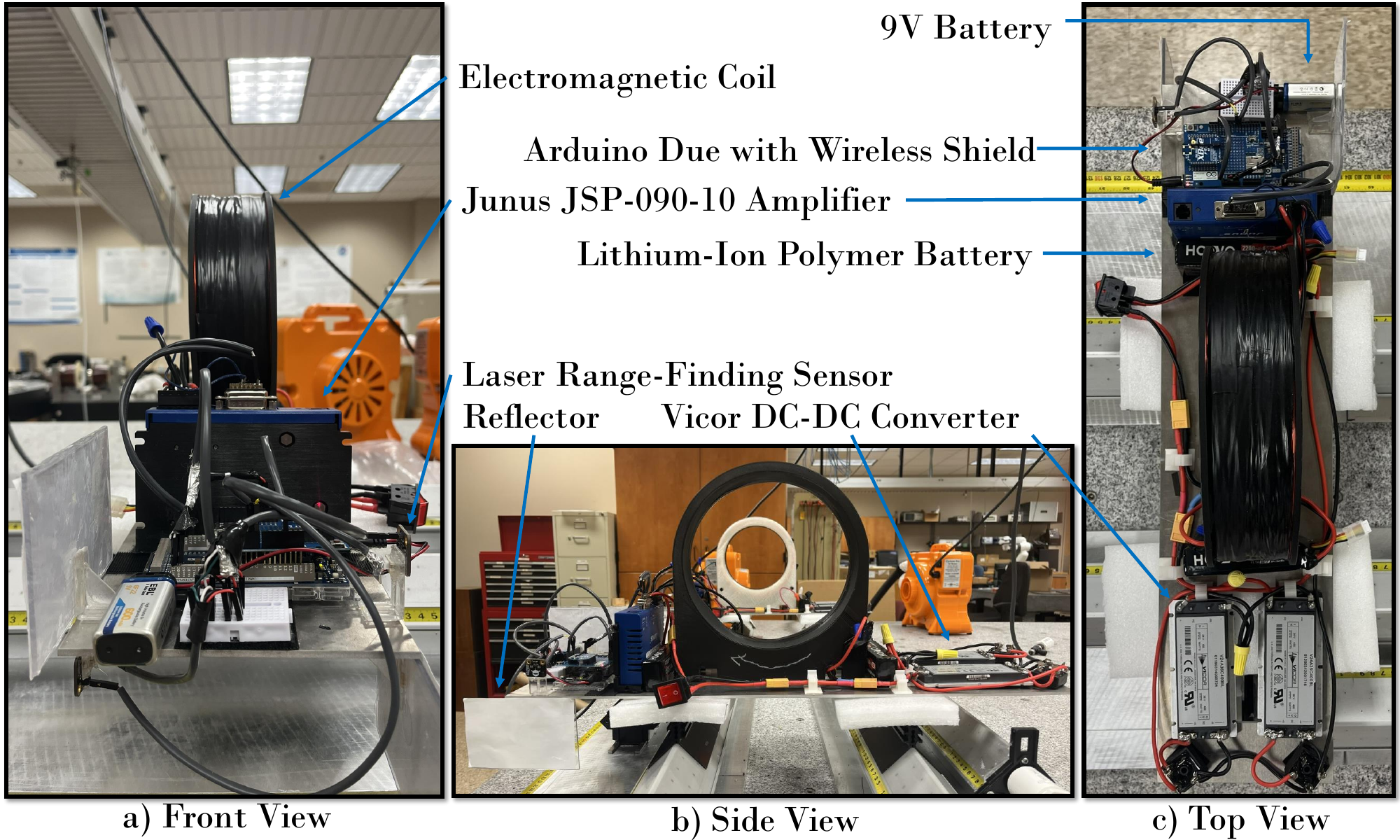}
    \caption{Front view, side view, and top view of the EAS unit on the air track, displaying the power electronics, the Arduino, and the laser range-finding sensor.}
    \label{fig:EAS_unit}
\end{figure}

{\subsection{Testbed Hardware}}
\label{sec:Exp_platform}

The experimental platform is shown in Fig.~\ref{fig:testbed}. The testbed includes four Cylone Pro air blowers, two Eisco PH0362A Linear Air Tracks, and three custom-designed one-dimensional electromagnetic actuation system (EAS) units. Each air track is attached to one Cylone Pro air blower. The air blowers supply the required air pressure to the linear air tracks. The EAS units sit on four air track gliders to allow for low-friction motion of the units on the air track. The air tracks are mounted on an aluminum jig-plate which is mounted on a flat and level surface table.

\begin{figure}[hbt!]
    \centering
    \includegraphics[width=0.5\textwidth,clip=true,trim= 0.3in 0.0in 0.5in 0in]{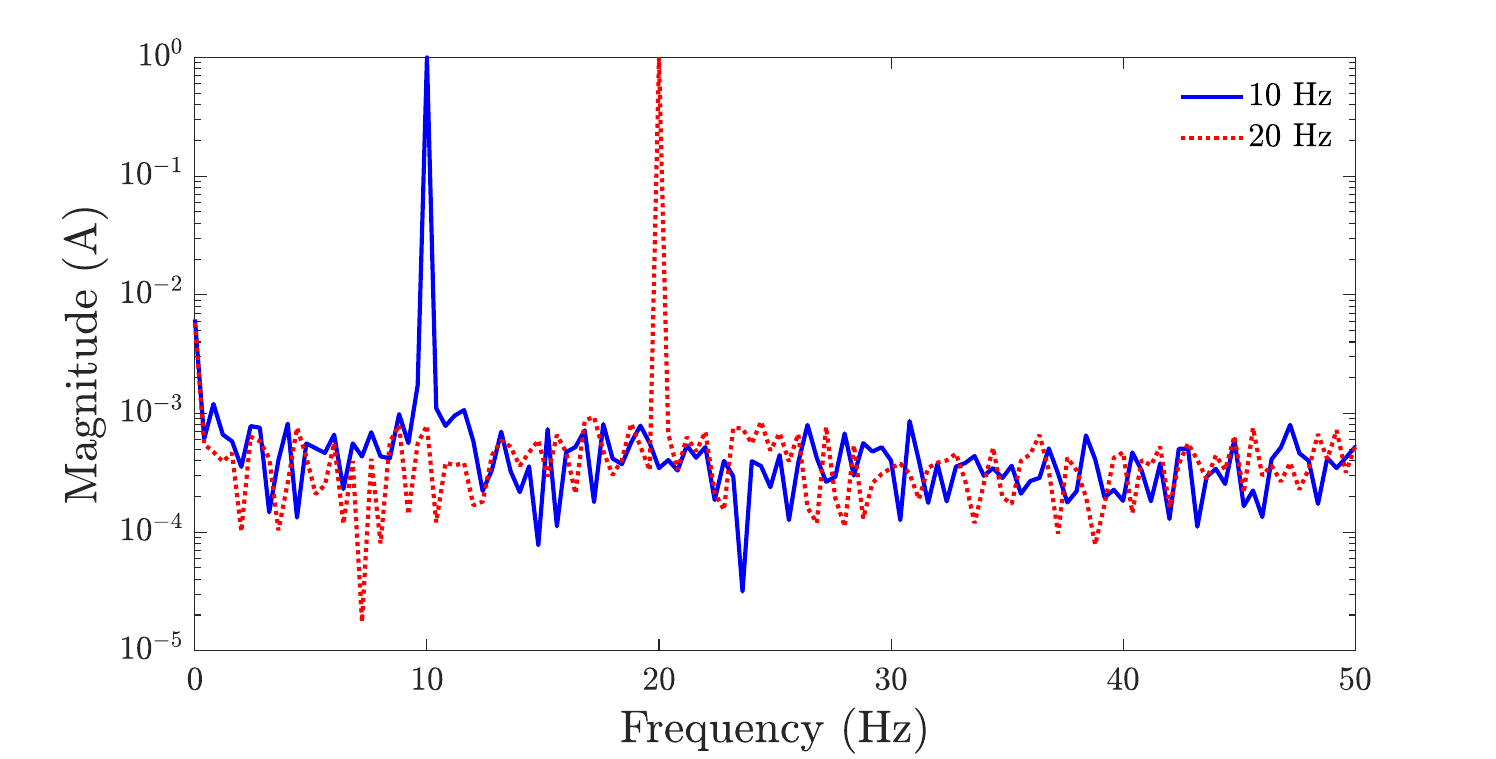}
    \caption{Frequency spectrum of amplifier current output with 10~Hz and 20~Hz sinusoidal command.}
    \label{fig:freq_specbench_tests}
\end{figure}

Each EAS unit has an electromagnetic coil and an electrical and electronics platform. Each electromagnetic coil has an area of $A= 0.1\pi^2$ m$^2$ and consists of $N= 500$ turns of 22 gauge copper wire. 
The inductance of each coil is 71 mH, resistance is 16 ohms, and the capacitance is below the measurement threshold (100 pF).
The electrical and electronics platforms include all components needed to supply power to the electromagnetic coils. 
The power electronics consist of 2 lithium-ion polymer batteries, 2 Vicor DC-DC converters, and a Copley Controls Junus JSP-090-10 amplifier, as shown in \Cref{fig:EAS_unit}.
Bench tests confirm that the amplifiers can produce sinusoidal current commands at the frequencies used in the experiments (i.e., 10~Hz, 20~Hz) without generating common harmonics that could compromise the accuracy of the time-averaged force calculation.
\Cref{fig:freq_specbench_tests} provides the power spectrum from bench tests, where the amplifier is commanded at 10~Hz and 20~Hz.
As shown, the amplifier is able to produce the desired commands without cross-coupling harmonics.

Each EAS unit has an Arduino Due microcontroller, which is responsible for data acquisition and feedback control. 
An XBee radio frequency (RF) module is connected to an Arduino wireless secure digital (SD) shield, which is attached to the microcontroller. 
A coordinator XBee RF module is also connected to a computer, enabling communication between the computer and the EAS units through a Digi XCTU application. 
The baud rate of the XBees are set at 115,200, and the data transmission frequency is 2.4 GHz. 
The satellites' Arduinos start the experiment simultaneously using a transmission of a 2.4 GHz synchronization signal from the XBee coordinator.
The control is updated with period $T$, which is the same for all satellites.
There was no significant time or frequency drift in the experiments; however, this could be a concern for long-duration maneuvers, necessitating periodic resynchronization of the satellites or adaptive frequency tuning.

The outer EAS units have a single laser ranging module (STM VL53L0X) connected to the microcontroller, while the middle EAS unit has 2 laser-ranging modules on either side. 
The laser ranging modules are used to measure the unit’s position relative to the other unit. 
The laser ranging module is mounted on an aluminum plate, which forms the base of each EAS unit. 
The laser ranging module faces a white reflector that is mounted to the other EAS unit.

Prior to each experiment, the laser ranging sensors are calibrated using a bias correction procedure, where the satellites are placed at known relative positions (between 0.3 m and 0.6 m), which are measured 1000 times. 
The mean bias from these 1000 measurements is subtracted from the data collected in that experiment. 
Initial measurement noise reduction is accomplished by using a 40-point average for the 2-satellite experiments, where 40 measurements are taken during each $T=0.1$~s period and averaged to obtain the recorded measurement for that period. 
A 20-point average is used for the 3-satellite experiments.

\subsection{1D Control Specialization}
\label{sec:1D_Control}

This section specializes to 1D, where all motion is in the $\mathbf{i}$ direction. 
Specifically, let $r_{ij,k} \in \mathbb{R}$, $v_{ij,k} \in \mathbb{R}$, and $d_{ij} \in \mathbb{R}$ be such that 
\begin{equation}
    \mathbf{r}_{ij,k} = r_{ij,k} \mathbf{i}, \quad \mathbf{v}_{ij,k} = v_{ij,k} \mathbf{i}, \quad \mathbf{d}_{ij} = d_{ij} \mathbf{i},  
\end{equation}
and it follows from \Cref{c_ij} that
\begin{equation}
    \mathbf{f}^*_{ij,k}= f^*_{ij,k} \mathbf{i}, 
    \label{eq:fijk_i}
\end{equation}
where the desired value for the intersatellite force function in the $\mathbf{i}$ direction is
\begin{equation}
    f^*_{ij,k} \triangleq -\frac{ 2 m |r_{ij,k}|^4 }{c_0} 
    \alpha_{ij} (( r_{ij,k} - d_{ij} ) + \beta v_{ij,k} ).
    \label{c_ij_1D}
\end{equation}

For implementation, we do not control magnetic moment $\Cref{u_i}$ directly. 
Instead, the input is the current $I_i :\left[0,\infty \right) \rightarrow \mathbb{R}$ in the $i$th coil, which is
\begin{equation}
    I_{i}(t) =  \sum_{{j} \in \mathcal{N}_i} I_{ij,k} \sin \omega_{ij}t,
    \label{eqn:I_i}
\end{equation}
where the current amplitudes are
\begin{equation}
    I_{ij,k} \triangleq I^*(r_{ij,k},f^*_{ij,k},i-j),
    \label{eqn:I_ijk}
\end{equation}
and
\begin{equation}
    I^*(r_{ij,k},f^*_{ij,k},i-j) \triangleq \begin{cases}
           -\dfrac{\text{sgn} (f^*_{ij,k})}{NA} \sqrt{  \dfrac{ |  f^*_{ij,k} | }{2}} \mathbf , &i-j<0, 
    \\
       \dfrac{\text{sgn}(r_{ij,k})}{NA} \sqrt{ \dfrac{ |  f^*_{ij,k} | }{ 2 } } \mathbf , &i-j>0 .
       \end{cases}
    \label{eqn:I_(r_f_i_j)}
\end{equation}
The current control \eqref{eqn:I_i} generates a piecewise sinusoidal magnetic moment with frequency $\omega_{ij}$ and amplitude
\begin{equation}
    p_{ij,k} = NA I_{ij,k}.
    \label{eqn:pijk_fn_NAI}
\end{equation}
Thus, \eqref{eqn:I_i} generates the magnetic moment $\Cref{u_i}$, where $\mathbf{p}_{ij,k}=p_{ij,k} \mathbf{i}$, which is equal to \Cref{eqn:p_ij_allocation} where $\mathbf{f}^*_{ij,k} = f^*_{ij,k} \mathbf{i}$ and $\mathbf{r}_{ij,k}=r_{ij,k} \mathbf{i}$.

To verify that $\mathbf{f}(\mathbf{r}_{ij,k}, \mathbf{p}_{ij,k}, \mathbf{p}_{ij,k})=\mathbf{f}^*_{ij,k}$, note that since $\mathbf{r}_{ij,k} = r_{ij,k} \mathbf{i}$, and $\mathbf{p}_{ij,k}=p_{ij,k} \mathbf{i}$, it follows from \eqref{f(r_ui_uj)} that
\begin{equation}
    \mathbf{f}(\mathbf{r}_{ij,k},\mathbf{p}_{ij,k},\mathbf{p}_{ji,k}) = f(r_{ij,k}, p_{ij,k}, p_{ji,k}) \mathbf{i},
    \label{eq:frp_frp_i}
\end{equation}
where
\begin{equation}
    f(r_{ij,k},p_{ij,k},p_{ji,k}) \triangleq -2 \sgn(r_{ij,k}) p_{ij,k} p_{ji,k}. \label{eq:f_sgn_r_p}
\end{equation}
Substituting \Cref{eqn:I_ijk,eqn:I_(r_f_i_j),eqn:pijk_fn_NAI} into \Cref{eq:f_sgn_r_p} yields $f(r_{ij,k},p_{ij,k},p_{ji,k})=f^*_{ij,k}$, which combined with \Cref{eq:fijk_i,eq:frp_frp_i} demonstrates that $\mathbf{f}(\mathbf{r}_{ij,k},\mathbf{p}_{ij,k},\mathbf{p}_{ji,k}) = \mathbf{f}^*_{ij,k}$.

\subsection{Control Modifications for Hardware Limitations}
\label{sec:modifications_current}

We introduce 2 modifications to the current control \Cref{c_ij_1D,eqn:I_i,eqn:I_ijk,eqn:I_(r_f_i_j)}. 
First, we introduce integral action to the desired force \Cref{c_ij_1D} to address steady-state error. 
Second, we introduce modifications to the current allocation \Cref{eqn:I_ijk} to address current amplitude saturation. 
This section redefines $f_{ij,k}^*$ and $I_{ij,k}$ given by \Cref{c_ij_1D,eqn:I_ijk} based on these modifications.

Let $\epsilon_0 > 0$ and $\epsilon_1 > \epsilon_0$, and let the integrator state $\xi_{ij,k}$ satisfy
\begin{equation}
    \xi_{ij,k} \triangleq \begin{cases}                    
                     \xi_{ij,k-1} + r_{ij,k}- d_{ij}  ,&  |r_{ij,k}  - d_{ij}| \in (\epsilon_0, \epsilon_1), 
                     \\
                     0   ,&   |r_{ij,k}  - d_{ij}| \notin (\epsilon_0, \epsilon_1),    \end{cases}      
                     \label{eq:xi_ij}
\end{equation}
where $\xi_{ij,0} \in \mathbb{R}$ is the initial condition. 
Note that \Cref{eq:xi_ij} is an integrator if the formation error $r_{ij,k} - d_{ij}$ has small magnitude (i.e., in the interval $(\epsilon_0,\epsilon_1)$). 
Otherwise, the integrator is disabled to prevent integrator windup during the transient phase when the position error can be relatively large.
Then, define the modified desired formation force 
\begin{equation}
    f^*_{ij,k} \triangleq -\frac{ 2 m |r_{ij,k}|^4 }{c_0} 
    \left( \alpha_{ij} \left( \left( r_{ij,k} - d_{ij} \right) + \beta v_{ij,k} \right) + \rho_{ij} \xi_{ij,k} \right). 
    \label{eq:f*_ij_1D}
\end{equation}
where $\rho_{ij}=\rho_{ji} \geq 0$.

Since the intersatellite force depends on the product of $I^*(r_{ij,k},f^*_{ij,k},i-j)$ and $ I^*(r_{ji,k},f^*_{ji,k},j-i)$, equal force can be achieved if one amplitude is increased and the other is decreased proportionally. 
Thus, we allocate more control authority to satellites with fewer neighbors. 
To do so, we multiply $I^*(r_{ij,k},f^*_{ij,k},i-j)$ with $\gamma_{ij}$ where $\gamma_{ij}>0$ and $\gamma_{ji}=1/\gamma_{ij}$.

Next, define
\begin{equation*}
    \bar{I}_{i,k} \triangleq \max_{t \in [kT,kT+T)}  \left| \sum_{j \in \mathcal{N}_i}  \gamma_{ij} I^*(r_{ij,k},f^*_{ij,k},i-j)  \sin \omega_{ij}t \right|,     
\end{equation*}
which is the maximum amplitude of the sum of sinusoidal currents over $t \in [kT,kT+T)$. 
Since the current is limited by the power capability of the EAS units, we modify $I_{ij,k}$ such that $\bar{I}_{i,k}$ does not exceed the maximum allowable current $\bar{I}>0$. 
The current implemented on the $i$th coil is \Cref{eqn:I_i}, where the current amplitude is
\begin{equation}
    I_{ij,k} \triangleq \begin{cases}                       
    \gamma_{ij} I^*(r_{ij,k},f^*_{ij,k},i-j)  ,&  \quad  \bar{I}_{i,k} \leq \bar{I},
        \\
    \gamma_{ij} \dfrac{ \bar{I} }{ \bar{I}_{i,k} }  I^*(r_{ij,k},f^*_{ij,k},i-j),&  \quad \bar{I}_{i,k} > \bar{I}.   \end{cases} 
    \label{eq:hat_I_ij}
\end{equation}
Then, it follows from \Cref{eqn:I_i,eq:hat_I_ij} that
 \begin{equation}
     \max_{t \in [kT,kT+T)} |I_i (t)| \leq \bar{I},  
 \end{equation}
which implies that the current implemented is limited to the maximum allowable current.

Next, we show that the resultant prescribed force is a scaling of the unconstrained prescribed force $f^*_{ij,k}$. 
Specifically, substituting \Cref{eqn:I_(r_f_i_j),eqn:pijk_fn_NAI,eq:hat_I_ij} into \Cref{eq:f_sgn_r_p} yields
\begin{equation*}
    f(r_{ij,k},p_{ij,k}, p_{ji,k}) =  \frac{\bar{I}^2}{(\max \{ \bar{I},\bar{I}_{i,k}\})( \max \{ \bar{I},\bar{I}_{j,k}\})} f^*_{ij,k}. 
    \label{eq:frc_curr_amp}
\end{equation*}
Furthermore, \Cref{avg_approx_F_ij,avg_approx_Fij_w_pij} imply that the approximate time-averaged intersatellite force is $\hat{\mathbf{F}}_{ij,k} = \hat{F}(r_{ij,k},p_{ij,k}, p_{ji,k}) \mathbf{i}$, where
\begin{equation}
    \hat{F}(r_{ij,k},p_{ij,k}, p_{ji,k}) \triangleq \frac{c_0}{2|r_{ij,k}|^4} f(r_{ij,k},p_{ij,k}, p_{ji,k}). 
\end{equation}


\subsection{Kalman Filter for State Estimation}
\label{sec:Kalman_filter}

Let $q_{ij,k}$ be the measurement of the relative position at time $kT$, which is taken on satellite $i$ using the onboard sensor. For each satellite, we design Kalman filters, where each Kalman filter uses $q_{ij,k}$ to obtain the estimates $\hat{r}_{ij,k}$ and $\hat{v}_{ij,k}$ of $r_{ij,k}$ and $v_{ij,k}$ for $j \in \mathcal{N}_i$.

It follows from \Cref{discrete_velocity,avg_approx_F_ij,avg_approx_Fij_w_pij} that $\hat{r}_{ij,k}$ approximately satisfies sampled-data double integrator dynamics where the input is 
\begin{align}
    \nu_{ij,k} &\triangleq \frac{1}{m} \left( \sum_{g \in \mathcal{N}_i } \hat{F}(\hat{r}_{ig,k},p_{ig,k}, p_{gi,k}) \right. \notag \\
    &\qquad - \left. \sum_{h \in \mathcal{N}_j }  \hat{F}(\hat{r}_{jh,k},p_{jh,k}, p_{hj,k}) \right).
    \label{input_rel_DI_1D}
\end{align}
Since satellite $i$ does not have information of all forces acting on satellite $j$, we consider only the intersatellite forces applied to satellite $j$ by the common neighbors of satellites $i$ and $j$. 
Thus, the estimate of $\nu_{ij,k}$ computed onboard satellite $i$ is 
\begin{align}
    \hat{\nu}_{ij,k} &\triangleq \frac{1}{m} \left( \sum_{g \in \mathcal{N}_i } \hat{F}(\hat{r}_{ig,k},p_{ig,k}, p_{gi,k})\right. \notag \\
    &\qquad - \left. \sum_{h \in \{i\} \cup  (\mathcal{N}_i \cap \mathcal{N}_j) } \hat{F}(\hat{r}_{jh,k},p_{jh,k}, p_{hj,k}) \right).
    \label{input_rel_DI_1D_app}
\end{align}
Let $V >0$ be the variance of the noise associated with the measurement $q_{ij,k}$. 
Then for all $k$, the filtered relative position $\hat{r}_{ij,k}$ and relative velocity $\hat{v}_{ij,k}$ are given by
\begin{equation}
    \begin{bmatrix}
        \hat{r}_{ij,k}\\
        \hat{v}_{ij,k}
    \end{bmatrix} = (A-LCA) \begin{bmatrix}
        \hat{r}_{ij,k-1}\\
        \hat{v}_{ij,k-1}
    \end{bmatrix} + (B-LCB) \hat{\nu}_{ij,k-1} + L q_{ij,k},
    \label{eqn:kalman_state_space}
\end{equation}
where 
\begin{equation}
    A \triangleq \begin{bmatrix}
        1 &T\\
        0 &1
    \end{bmatrix}, \quad B \triangleq \begin{bmatrix}
        0.5 T^2\\
        T
    \end{bmatrix}, \quad C \triangleq \begin{bmatrix}
        1 &0
    \end{bmatrix}, 
    \label{kalman_matrices)}
\end{equation}
and 
\begin{equation}
L \triangleq P C^{\text{T}} (CP C^{\text{T}} + V)^{-1}, 
\label{eqn:kalman_gain}
\end{equation}
where $P$ is the solution to the discrete algebraic Riccati equation 
\begin{equation}
    A^{\text{T}} P A - P - A^{\text{T}} P C (C^{\text{T}} P C + V)^{-1} C^{\text{T}} P A + W=0, \label{eq:dis_alg_ric}
\end{equation}
where $W \triangleq w BB^{\mathrm{T}}$ and $w>0$ is the variance of the disturbance force. 
The performance of this filter may degrade in the presence of non-Gaussian disturbances (e.g., disturbances from the airflow leaving the air tracks in the experiments, atmospheric drag in low Earth orbit for on-orbit deployment). 
In this case, other state estimation methods (e.g., particle filters) could be considered.

\section{Experimental Results and Discussion}
\label{sec:Exp_Results_discussion}

We present results from open-loop and closed-loop experiments using the EAS units described in \Cref{sec:Exp_platform} with the control algorithm presented in \Cref{sec:1D_Control} with the modifications presented in \Cref{sec:modifications_current}. 
The control update period $T=0.1$~s is selected based on a tradeoff between a variety of factors, including the time-averaged force approximation, speed of the closed-loop response, computational capability of the microprocessor, and response time of the electromagnetic dynamics.
First, recall that the approximate time-averaged formation dynamics (i.e., \eqref{discrete_velocity} with $\bar{\mathbf{F}}_{ij} (k)=\hat{\mathbf{F}}_{ij,k}$) is developed under the assumption that $\mathbf{r}_{ij}$ does not change significantly over period $T$.
Thus, $T$ is selected to be small enough that this assumption is reasonable given the anticipated speed of the closed-loop dynamics. 
Conversely, $T=0.1$~s is large enough such that all data acquisition and computations can be performed in real time onboard the microprocessor. 
Moreover, $T=0.1$~s is approximately 23 times larger than 4.4~ms time constant of the electromagnetic dynamics, which implies that $T$ is large enough for those dynamics to be neglected.

Using experimental data, the noise variances are estimated as $V=1.2 \times 10^{-6}$ $\mathrm{m}^2$ and $w= 5 \times 10^{-6}$~$\mathrm{m}^2/\mathrm{s}^4$ for the 2-satellite experiments, and $V=2 \times 10^{-6}$ $\mathrm{m}^2$ and $w= 5 \times 10^{-6}$~$\mathrm{m}^2/\mathrm{s}^4$ for the 3-satellite experiments. 
Then, \Cref{kalman_matrices),eqn:kalman_gain,eq:dis_alg_ric} are used to obtain $P$ and $L$ provided in \Cref{table:kalman_parameters}. 
The Kalman filter is initialized with $\hat{r}_{ij,0}=q_{ij,0}$ and $\hat{v}_{ij,0}=0$~m/s.

The plots of the experiments include $|q_{ij,k}|$, $|\hat{r}_{ij,k}|$, $\hat{v}_{ij,k}$, $I_{ij,k}$, and $\hat{F}(r_{ij,k},p_{ij,k},p_{ji,k})$. 
\Cref{table:data_measurements} shows the range of data measured or computed onboard the satellites during the experiments.

\begin{table}[h!]
\caption{Kalman Filter Parameters for 2- and 3-Satellite Experiments}
\begin{center}
 \scriptsize
\resizebox{\columnwidth}{!}{%
\begin{tabular}{c|c|c}
 & 2 Satellite & 3 Satellite \\
\hline
$V$ (m$^2$)   &$1.2 \times 10^{-6}$ & $2 \times 10^{-6}$ \\[4pt] 
$w$ (m$^2$/s$^4$)  &$5 \times 10^{-6}$ & $5 \times 10^{-6}$   \\[4pt]
$P$   &$\begin{bmatrix}
        0.2686    &0.2710\\
    0.2710    &0.5206\\
    \end{bmatrix} \times 10^{-6}$   & $\begin{bmatrix}
0.3891    &0.3456\\
    0.3456    &0.5879
\end{bmatrix} \times 10^{-6}$   \\[6pt]
$L$   &$\begin{bmatrix}
        0.2013 &0.1845
    \end{bmatrix}^{\mathrm{T}}$   & $\begin{bmatrix} 0.1773 & 0.1447 \end{bmatrix}^{\mathrm{T}}$
\end{tabular}%
}
\end{center}
\label{table:kalman_parameters}
\end{table}

\begin{table}[h!]
{
\caption{Range of Data Measured or Computed Onboard Satellites During Experiments}
\begin{center}
 \scriptsize
\resizebox{\columnwidth}{!}{%
\begin{tabular}{c|c|c}
 & 2 Satellite & 3 Satellite \\
\hline
$|q_{ij}|$ (m)   &[0.38,0.54] & [0.34,0.46] \\
[2pt] 
$|\hat{r}_{ij}|$ (m)   &[0.39,0.53] &[0.34,0.46]    \\
[2pt]
$\hat{v}_{ij}$ ($10^{-2}$ m/s)   &[-1.1,0.97]   &[-1.1,1.1]    \\[2pt]
$I_{ij}$ (A)   &[-2.35,1.44]  & [-2.15,2.08]\\
[2pt]
$\hat{F}$ ($10^{-3}$ N)  &[-6.6,5.7]  & [-4.9,4.9]
\end{tabular}%
}
\end{center}
\label{table:data_measurements}
}
\end{table}


\subsection{Two-Satellite Validation in Open and Closed Loop}
\label{sec:Exp_Results_2_sat}

We present 2 open-loop experiments and 3 closed-loop experiments with 2 satellites. 
The left satellite is numbered~1 and the right satellite is numbered~2. 
The actuation frequency is $\omega_{12}= \omega_{21} = 40 \pi$ rad/s.
First, we present open-loop experiments to demonstrate that the experimental testbed is capable of generating attractive and repulsive forces.

\begin{figure}[ht!]
    \centering
    \includegraphics[width=0.50\textwidth,clip=true,trim= 0.0in 0.6in 0in 0.8in]{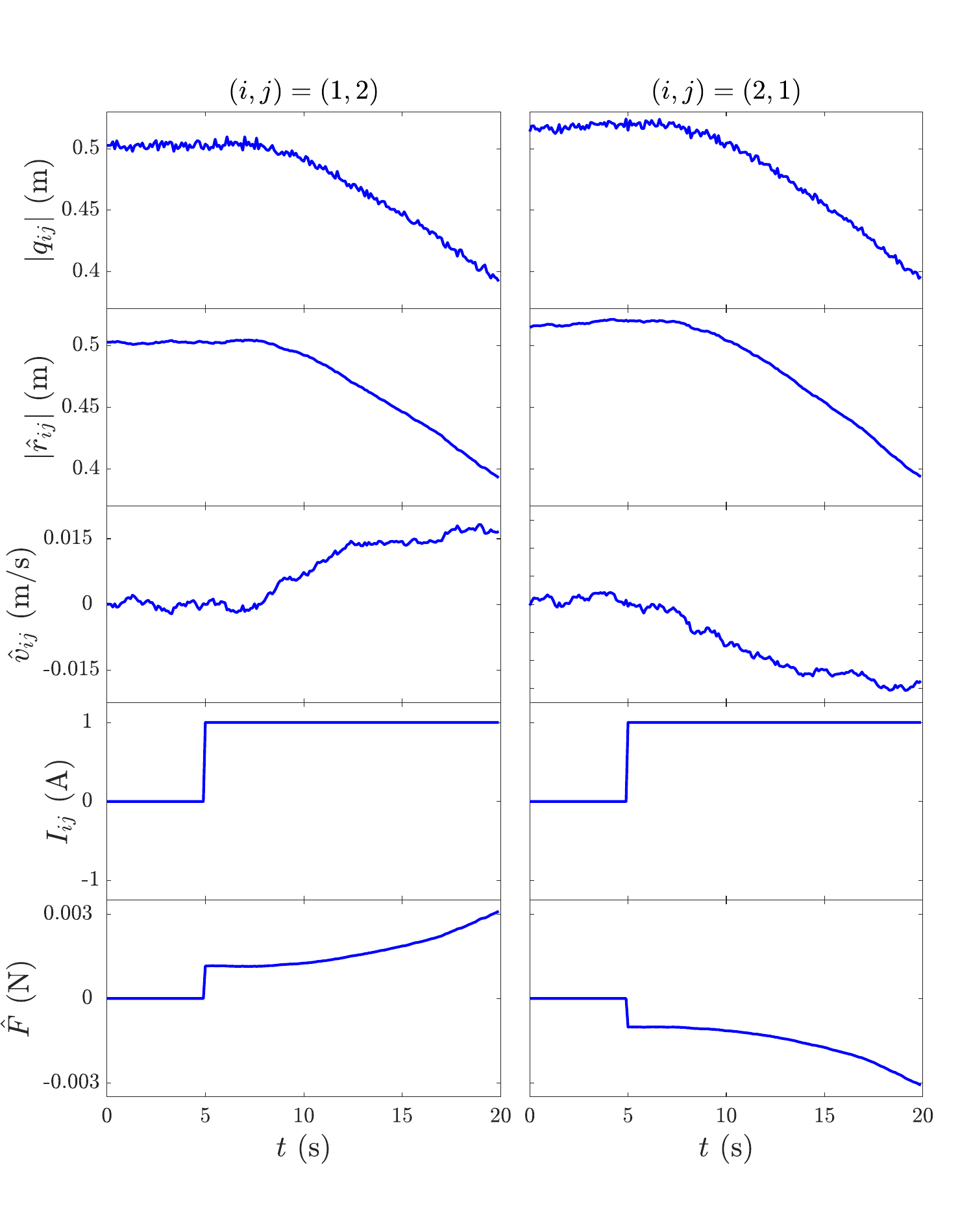}
    \caption{Open-loop attraction between 2 satellites where $I_{12,k}=1$~A and $I_{21,k}=1$~A for $t>5$ seconds. 
    The sinusoidal currents on the satellites are in phase, which results in an attractive force.}
    \label{fig:ol_attraction_2}
\end{figure}

\begin{experiment}\rm 
This experiment demonstrates open-loop attraction. \Cref{fig:ol_attraction_2}  shows the experimental results, where the first row shows the raw relative-position measurements $|q_{ij}|$, the second and third rows show the filtered measurements $|\hat{r}_{ij}|$ and $\hat{v}_{ij}$, the fourth row shows the current $I_{ij}$, and the fifth row shows the time-averaged intersatellite force $\hat{F}(r_{ij,k},p_{ij,k}, p_{ji,k})$. 
The first column shows the data measured and computed onboard satellite~1, and the second column shows the data measured and computed onboard satellite~2.

The control is turned on at $t=5$~s with amplitudes $I_{12}=1$~A and $I_{21}=1$~A (sinusoidal currents are in phase). 
Then, $|\hat{r}_{ij}|$ decreases from 0.508~m to 0.38~m over the next 15~s, which demonstrates attraction. 
There is a slight offset between the initial estimates $|\hat{r}_{12,0}|$ and $|\hat{r}_{21,0}|$, which resolves at smaller distance. 
This can be explained by sensor and reflector misalignment, and laser-ranging sensor bias at larger distances. 
\exampletriangle
\end{experiment}

%

\begin{figure}[ht!]
    \centering
    \includegraphics[width=0.50\textwidth,clip=true,trim= 0.0in 0.6in 0in 0.8in]{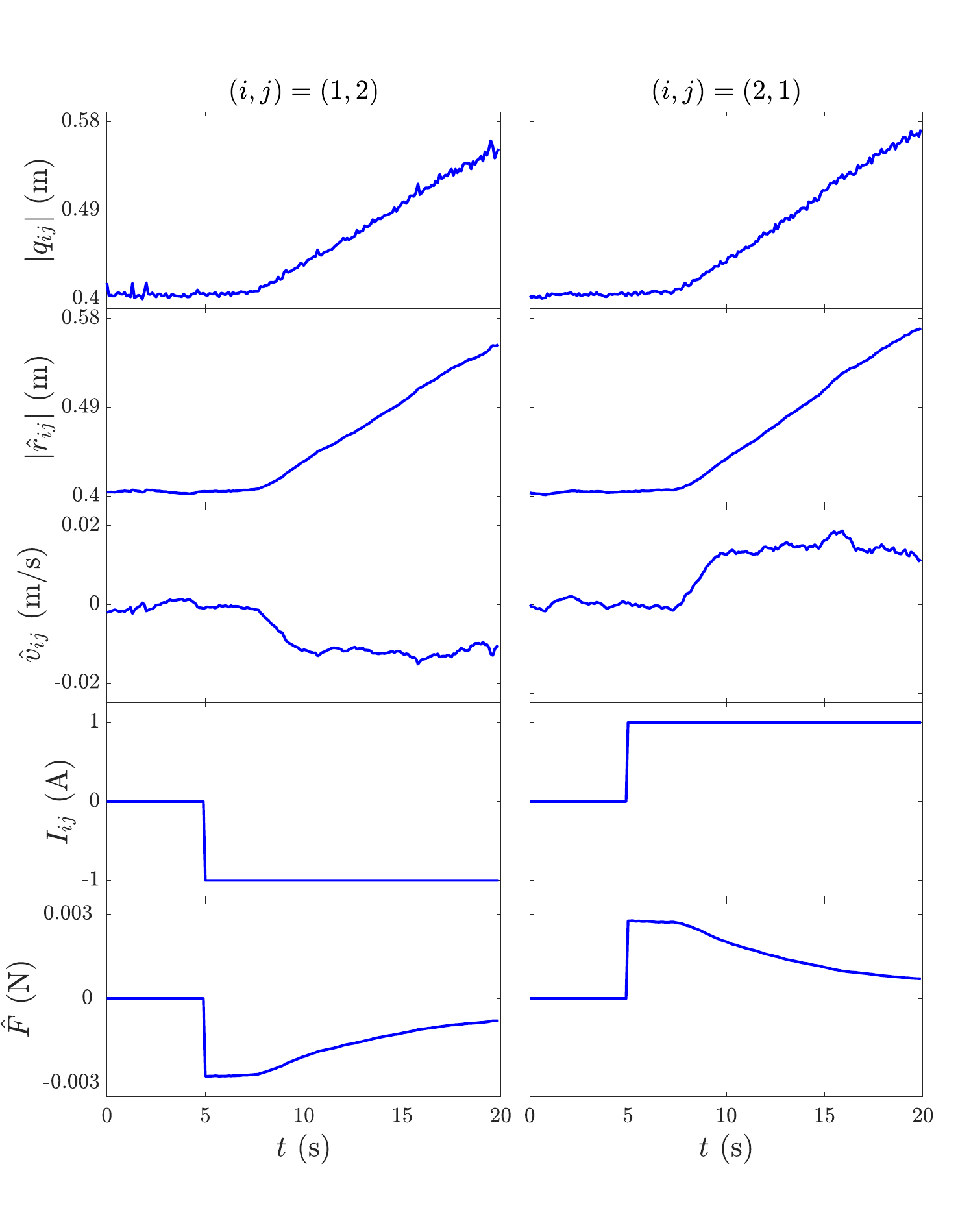}
    \caption{Open-loop repulsion between 2 satellites where $I_{12,k}=-1$~A and $I_{21,k}=1$~A for $t>5$ seconds. 
    The sinusoidal currents on the satellites are out of phase, which results in a repulsive force.}
    \label{fig:ol_repulsion_2}
\end{figure}

\begin{experiment}\rm 
This experiment demonstrates open-loop repulsion. \Cref{fig:ol_repulsion_2} shows that the control is turned on at $t=5$~s with amplitudes $I_{12}=-1$~A and $I_{21}=1$~A (sinusoidal currents are out of phase). 
Then, $|\hat{r}_{ij}|$ increases from 0.404~m to 0.56~m over the next 15~s, which demonstrates repulsion.   
\exampletriangle
\end{experiment}

Next, we present experiments demonstrating closed-loop control. 
The control $(I_{12,k}, I_{21,k})$ is implemented with $\alpha_{ij}=0.0158$, $\beta=6.89$, $\rho_{ij}=0$, and $\bar{I}=2.35$ A. 
Note that $\alpha_{ij}$ and $\beta$ are selected such that $\alpha_{ij} m/(c_0 N^2 A^2)=812$ and $\alpha_{ij} \beta m/(c_0 N^2 A^2)=5600$. 
These gains are selected based on a tradeoff between response characteristics and hardware capabilities (e.g., current limits, sensor noise). 
Notably, the noise in the relative velocity estimate $\hat v_{ij}$ limits the magnitude of the feedback gains. 
This limit could potentially be mitigated by using different sensors. 
Initial gains were selected such that the idealized analytic closed-loop double-integrator dynamics have zero overshoot and less than 20~s settling time (i.e., time it takes $\hat{r}_{ij}$ to reach and stay within 0.0075~m or approximately 2\% of its final value).
The experimental response with these gains is too aggressive, amplifying noise in the velocity-feedback term and saturating the current.
Next, $\beta$ was decreased to reduce noise amplification in the velocity-feedback term and limit current saturation. 
Finally, $\alpha_{ij}$ was increased to recover experimental settling time less than 25~s and overshoot less than 0.02~m. 
The final values yield settling time of approximately 20~s and overshoot less than 0.01~m with the idealized analytic closed-loop double-integrator dynamics.

\begin{figure*}[!t]
    \begin{subfigure}{0.49\textwidth}
        \centering
        \includegraphics[width=1\textwidth,clip=true,trim= 0.0in 0.25in 0in 0.3in]{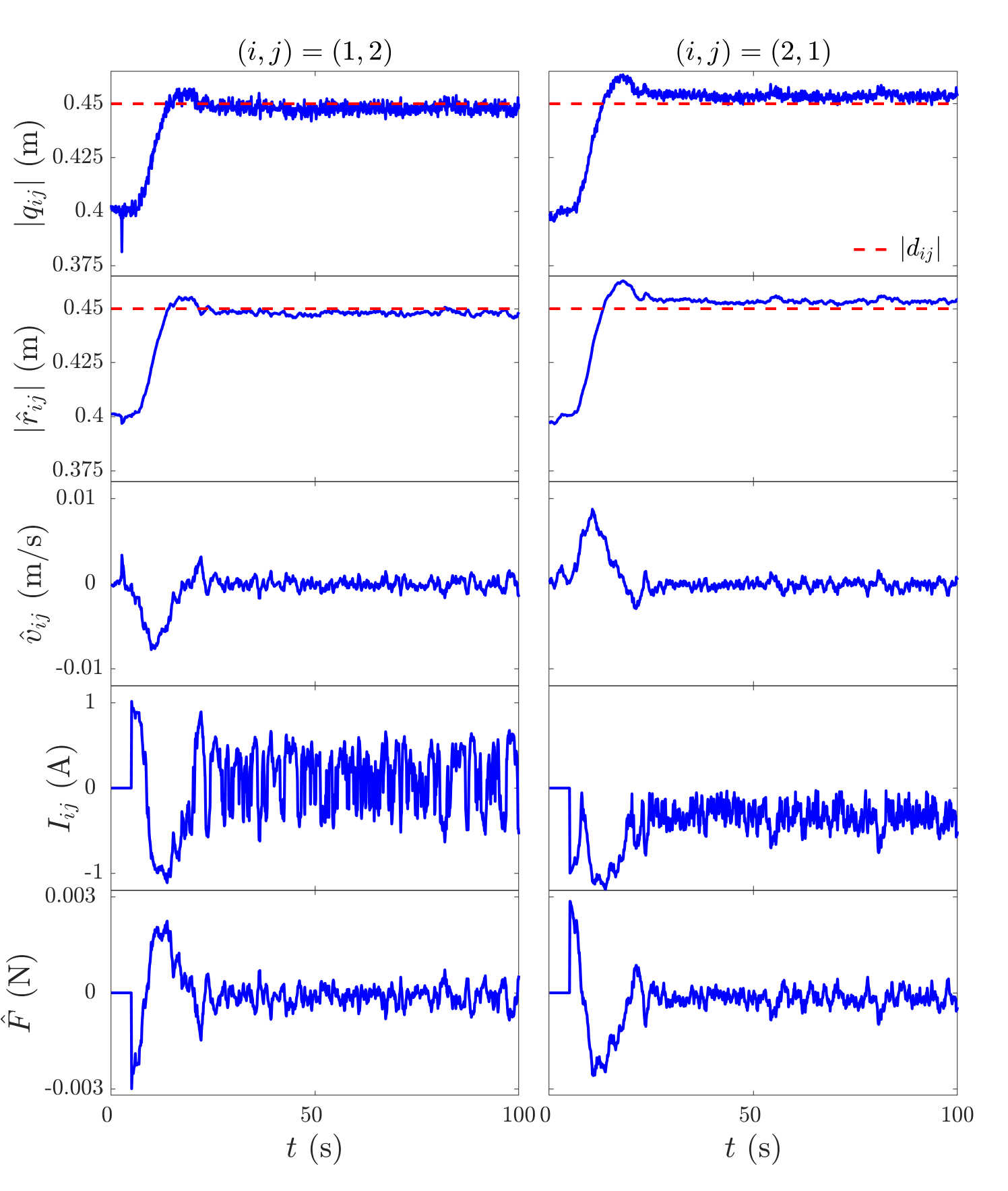}
        \caption{Experiment}
        \label{fig:cl_repulsion_2}
    \end{subfigure}
    \hfill
    \begin{subfigure}{0.49\textwidth}
        \centering
        \includegraphics[width=1\textwidth,clip=true,trim= 0.0in 0.25in 0in 0.3in]{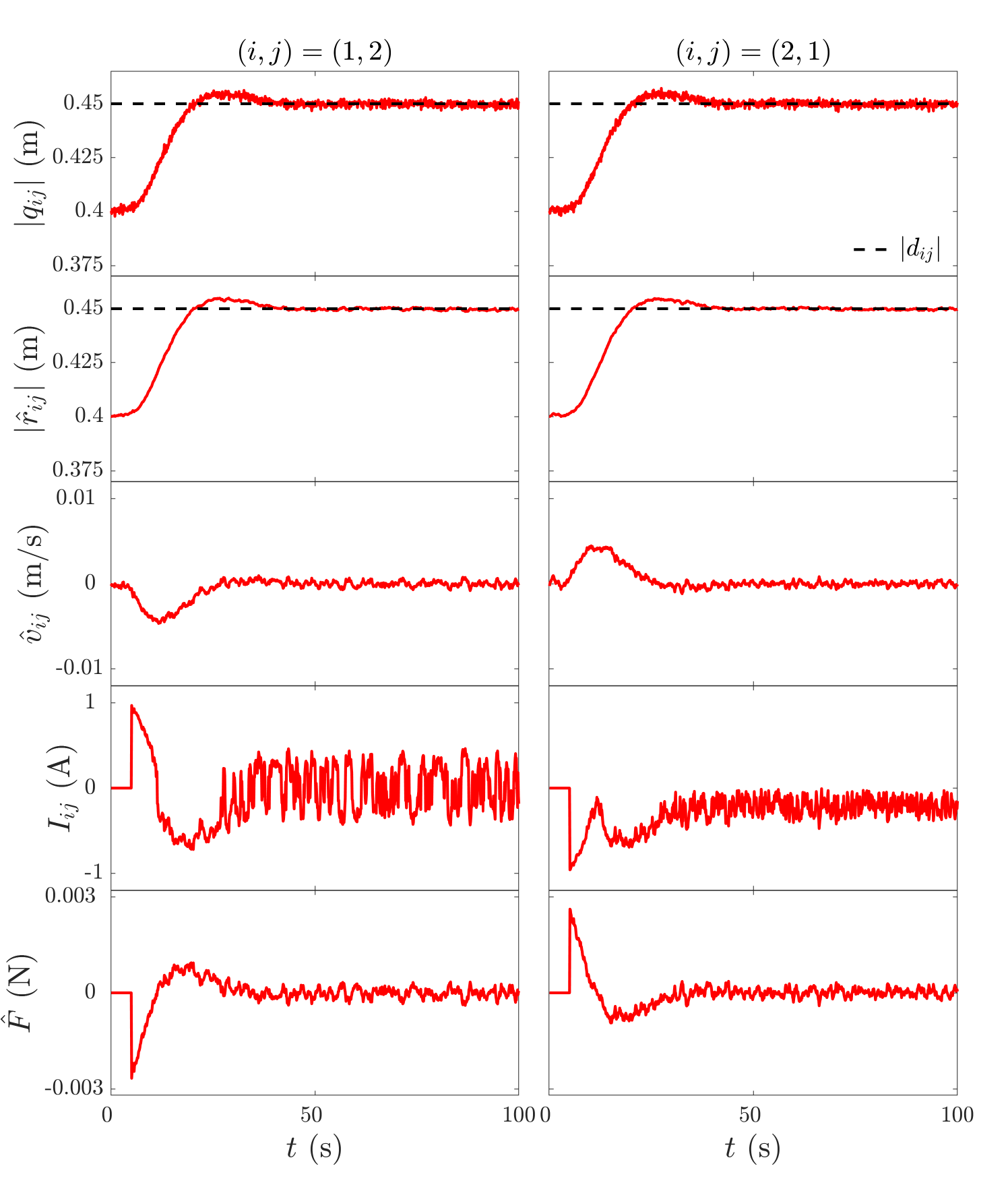}
        \caption{Simulation}
        \label{fig:cl_repulsion_sim_2}
    \end{subfigure}
    \caption{Closed-loop repulsion with 2 satellites. 
    Satellites start at approximately 0.4~m apart, and the desired relative position is $|d_{ij}|=0.45$~m.}
\end{figure*}

\begin{experiment}\rm 
\label{exp:2_sat_repulsion} 
This experiment demonstrates closed-loop repulsion. The satellites start at a relative distance of approximately 0.4~m. 
The objective is to reach a desired relative distance $|d_{ij}|=0.45$~m. 
\Cref{fig:cl_repulsion_2} shows that $|\hat{r}_{ij}|$ settles to $|d_{ij}|=0.45$~m by approximately 20~s with low steady-state error (SSE)
Specifically, the settling time is $19$~s.
The mean SSE (i.e., mean $| \hat{r}_{ij}-d_{ij} |$ over the last 30~s) is $7.7 \times 10^{-4}$~m.
Note that there is a sharp decrease in the measurement $|q_{12}|$ at $t=2.5$~s, which is due to a sensor dropout.
The variance of the unfiltered SSE $ q_{ij}-d_{ij} $ is $3.1\times 10^{-6}$~m$^2$, where $1.2\times 10^{-6}$~m$^2$ is attributed to sensor noise.
The remaining variance is potentially due in part to unmodeled external disturbances (e.g., air disturbance from the tracks) as well as unmodeled dynamics (e.g., friction, near-field electromagnetic forces), which can influence SSE through feedback.

Next, we compare the experimental results with simulations of the closed-loop \Cref{accel_i,u_i} where $\mathbf{p}_{ij,k}=p_{ij,k} \mathbf{i}$. 
We initialize the simulation using the same initial conditions as in the experiment. 
The simulated relative position includes zero-mean Gaussian-white sensor noise with variance $V=1.2\times10^{-6}$~m$^2$, which is the sensor noise variance estimated from experimental data. 
Then, the filtered relative position and velocity are obtained from the Kalman filter in \Cref{sec:Kalman_filter}, which is the method used in the experiment.
\Cref{fig:cl_repulsion_sim_2} shows the simulation results, which qualitatively match the experimental results in \Cref{fig:cl_repulsion_2}.

To compare the experiment and simulation quantitatively, we use 4 metrics: overshoot $\hat{r}_{\mathrm{os}}$; settling time $T_{\mathrm{s}}$; maximum value of $|\hat{F}(r_{ij,k},p_{ij,k}, p_{ji,k})|$; and 
\begin{equation}
    P_{\hat F} \triangleq \sqrt{\frac{1}{N_{\mathrm{t}}} \sum_{k=1}^{N_{\mathrm{t}}} \hat{F}(r_{ij,k},p_{ij,k}, p_{ji,k})^2}, 
\end{equation}
which is the root mean square of the force, where $N_{\mathrm{t}}$ is the number of data points. 
\Cref{Table:2_sat_experiment_vs_simulation} presents these metrics for the experiment and the simulation. 
The experiment and the simulation have similar $T_{\mathrm{s}}$ and $\max|\hat{F}|$. 
The difference in these metrics between the experiment and simulation is less than 15\%.
The experiment has 70\% more overshoot than the simulation; however, the overshoot is small (less than 1 cm) for both. 
The root mean square of the force in the experiment is 51\% larger than in the simulation. 
These differences can be partly explained by external disturbances present in the experiment that are not included in the simulation. 
For example, the air tracks are not perfectly level over the entire length; the airflow from the air tracks creates disturbances forces; and the initial velocity in the experiment is not perfectly known, so there is mismatch between this condition in the experiment and the simulation. 
Differences can also be partly explained by near-field electromagnetic effects \cite{Schweighart2005Th} that are not included in the simulation.
\exampletriangle
\end{experiment}

\begin{table}[hbt]
\centering
\scriptsize
\caption{Metrics for \Cref{exp:2_sat_repulsion} and Simulation}
\begin{center}
\begin{tabularx}{\columnwidth}{c|YY|YY}
\multirow{2}{*}{\ } & \multicolumn{2}{c|}{\Cref{exp:2_sat_repulsion}} & \multicolumn{2}{c}{Simulation} \\
\cline{2-5}
 &$(i,j)=(1,2)$  &$(i,j)=(2,1)$  &$(i,j)=(1,2)$  &$(i,j)=(2,1)$ \\
\hline
$\hat{r}_{\mathrm{os}}$ (cm) &0.76	&0.95 &0.55	&0.47\\
$T_{\mathrm{s}}$ (s) &18.7  &19.1   &17.2   &17.4\\
$\max |\hat{F}|$ (N) &$2.99 \times 10^{-3}$   &$2.85 \times 10^{-3}$  &$2.67 \times 10^{-3}$  &$2.61 \times 10^{-3}$\\
$P_{\hat F}$ (N) &$2.02\times 10^{-4}$   &$2.27\times 10^{-4}$   &$1.43\times 10^{-4}$   &$1.41\times 10^{-4}$\\
\end{tabularx}
\end{center}
\label{Table:2_sat_experiment_vs_simulation}
\end{table}

%

\begin{figure*}[!t]
    \begin{subfigure}{0.49\textwidth}
        \centering
        \includegraphics[width=1\textwidth,clip=true,trim= 0.0in 0.25in 0in 0.3in]{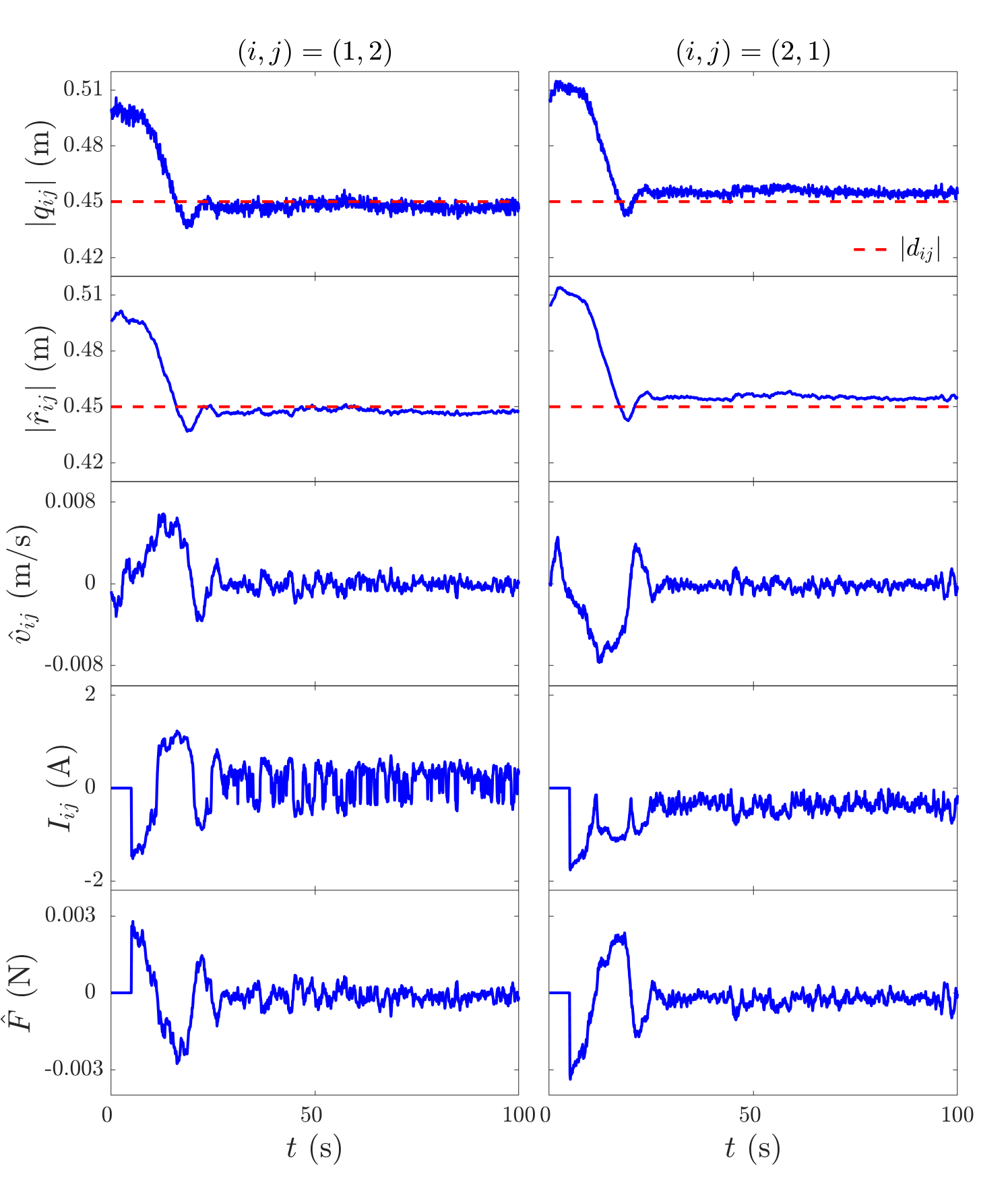}
        \caption{Experiment}
        \label{fig:cl_attraction_2}
    \end{subfigure}
    \hfill
    \begin{subfigure}{0.49\textwidth}
        \centering
        \includegraphics[width=1\textwidth,clip=true,trim= 0.0in 0.25in 0in 0.3in]{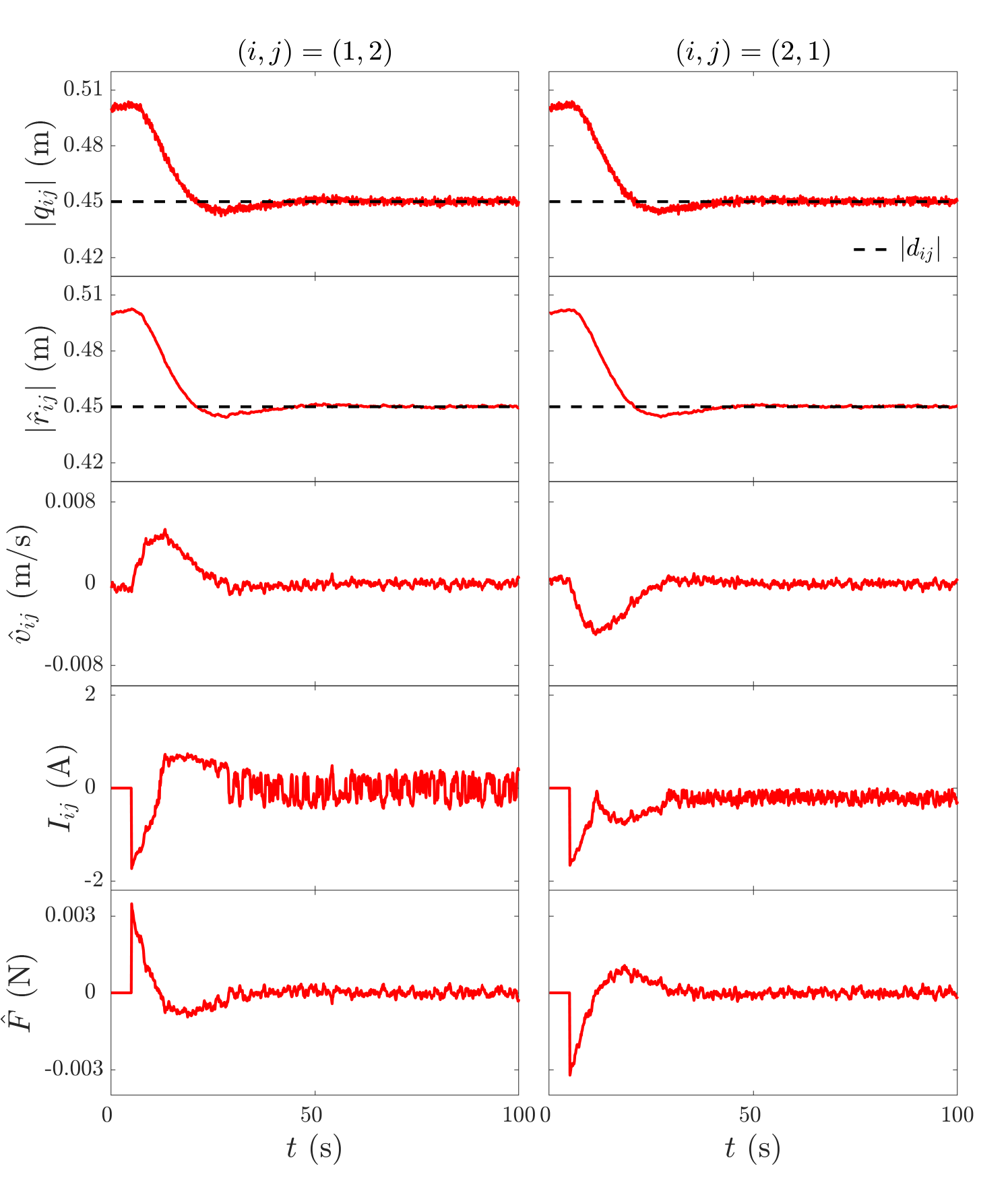}
        \caption{Simulation}
        \label{fig:cl_attraction_sim_2}
    \end{subfigure}
    \caption{Closed-loop attraction with 2 satellites.
    Satellites start at approximately 0.5~m apart, and the desired relative position is $|d_{ij}|=0.45$~m.}
\end{figure*}

\begin{experiment}\rm 
\label{exp:2_sat_attraction} 
This experiment demonstrates closed-loop attraction. 
The satellites start at a relative distance of approximately $0.5$~m. 
The objective is to reach a desired relative distance $|d_{ij}|=0.45$~m. 
\Cref{fig:cl_attraction_2} shows that $|\hat{r}_{ij}|$ settles to $|d_{ij}|=0.45$~m at 20~s. 
The mean SSE $| \hat{r}_{ij}-d_{ij} |$ over the last 30~s is $5.3 \times 10^{-4}$~m.
Note that the satellites move before the control is turned on at $t=5$~s. 
This can be explained by a slight impulse introduced at the start of the experiment and airflow from the air tracks creating disturbances. 
The variance of the unfiltered SSE $ q_{ij}-d_{ij} $ is $2.9\times 10^{-6}$~m$^2$, which is comparable to \Cref{exp:2_sat_repulsion}.

\Cref{fig:cl_attraction_sim_2} shows the simulation results, which qualitatively matches the experimental results in \Cref{fig:cl_attraction_2}. 
\Cref{Table:2_sat_experiment_attraction} presents the metrics $\hat{r}_{\mathrm{os}}$, $T_{\mathrm{s}}$, $\max|\hat{F}|$, and $P_\rmF$. 
The quantitative comparison is similar to \Cref{exp:2_sat_repulsion}. 
In particular, $T_{\mathrm{s}}$ and $\max|\hat{F}|$ are comparable between the experiment and the simulation. 
The difference in these metrics between the experiment and simulation is less than 15\%.
In contrast, $\hat{r}_{\rm{os}}$ and $P_\rmF$ are larger in the experiment. 
These differences can be attributed to the causes discussed in \Cref{exp:2_sat_repulsion}. 
\exampletriangle
\end{experiment}

\begin{table}[hbt]
\centering
\scriptsize
\caption{Metrics for \Cref{exp:2_sat_attraction} and Simulation}
\begin{center}
\begin{tabularx}{\columnwidth}{c|YY|YY}
\multirow{2}{*}{\ } & \multicolumn{2}{c|}{\Cref{exp:2_sat_attraction}} & \multicolumn{2}{c}{Simulation} \\
\cline{2-5}
 &$(i,j)=(1,2)$  &$(i,j)=(2,1)$  &$(i,j)=(1,2)$  &$(i,j)=(2,1)$ \\
\hline
$\hat{r}_{\mathrm{os}}$ (cm) &1.02	&1.22 &0.59	&0.58\\
$T_{\mathrm{s}}$ (s) &20.2   &20.5   &17.9   &17.9\\
$\max |\hat{F}|$ (N) &$2.79 \times 10^{-3}$   &$3.38 \times 10^{-3}$  &$3.48 \times 10^{-3}$  &$3.22 \times 10^{-3}$\\
$P_{\hat F}$ (N) &$2.38\times 10^{-4}$   &$2.62\times 10^{-4}$   &$1.66\times 10^{-4}$   &$1.64\times 10^{-4}$\\
\end{tabularx}
\end{center}
\label{Table:2_sat_experiment_attraction}
\end{table}

\begin{experiment}\rm \label{exp:multi_set_point} 
This experiment demonstrates multiple maneuvers in succession. 
The satellites start at a relative distance of approximately 0.50~m. 
The initial objective is to reach a desired relative distance $|d_{ij}|=0.4$~m. 
Next, the desired relative position is switched to $0.5$~m at $t=60$~s. 
\Cref{fig:cl_multiset_2} shows that $|\hat{r}_{ij}|$ settles to $|d_{ij}|=0.40$~m at approximately 35~s, and $|\hat{r}_{ij}|$ settles to $|d_{ij}|=0.50$~m by approximately 90~s, which demonstrates that the satellites can achieve multiple formation maneuvers.

\begin{figure*}[!t]
    \begin{subfigure}{0.49\textwidth}
        \centering
        \includegraphics[width=1\textwidth,clip=true,trim= 0.0in 0.25in 0in 0.3in]{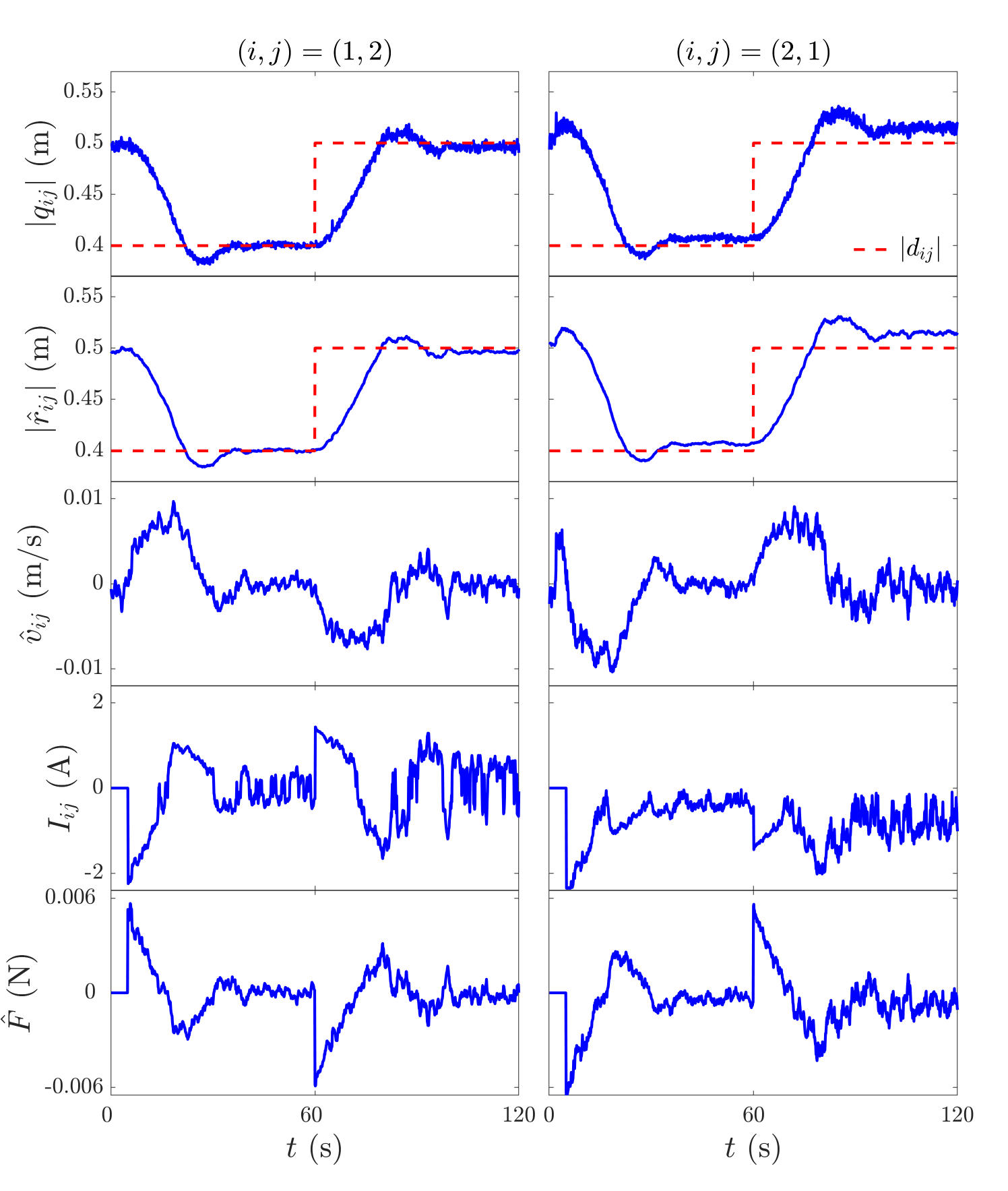}
        \caption{Experiment}
        \label{fig:cl_multiset_2}
    \end{subfigure}
    \hfill
    \begin{subfigure}{0.49\textwidth}
        \centering
        \includegraphics[width=1\textwidth,clip=true,trim= 0.0in 0.25in 0in 0.3in]{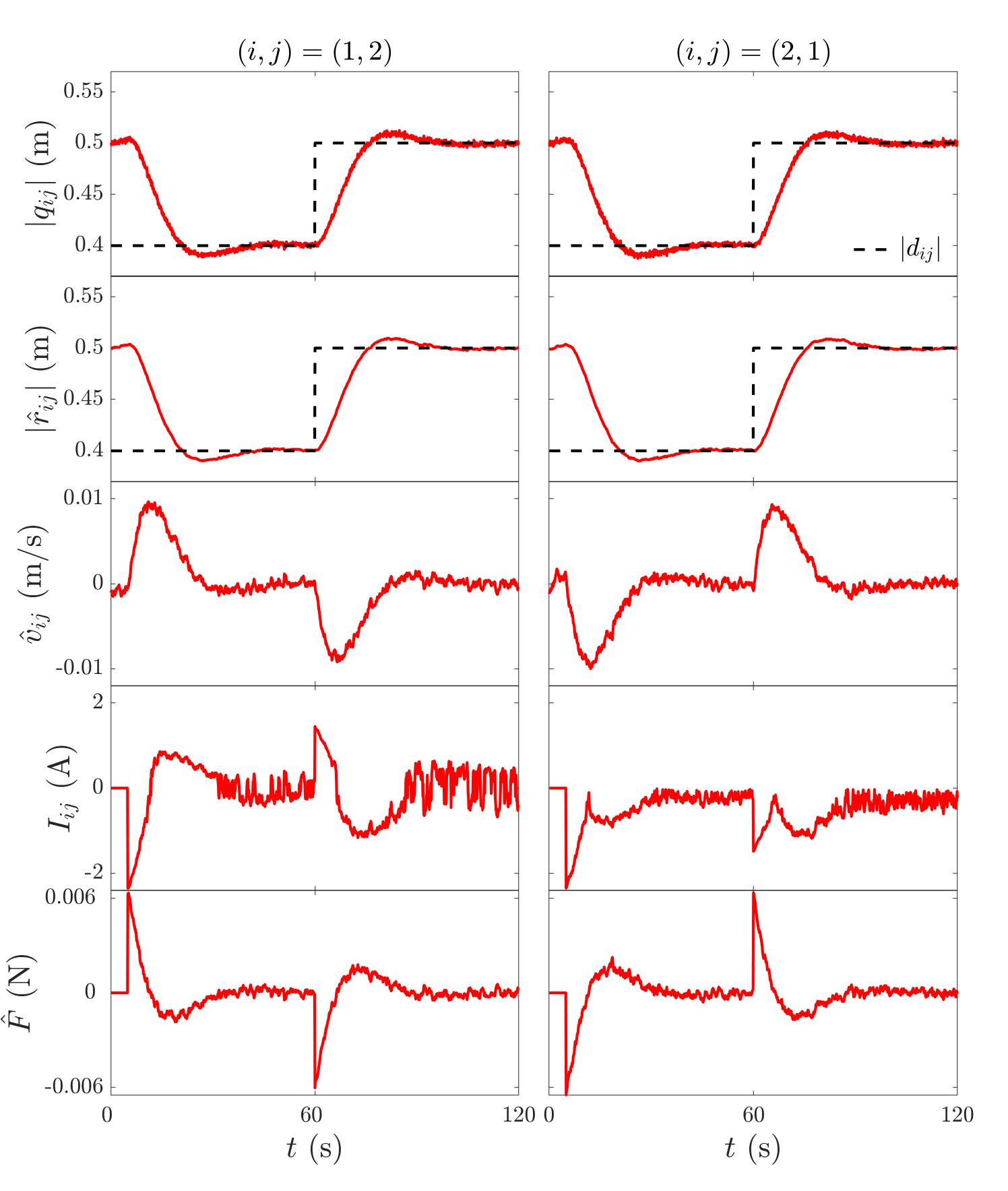}
        \caption{Simulation}
        \label{fig:cl_multiset_sim_2}
    \end{subfigure}
    \caption{Multiple maneuvers with 2 satellites, where the satellites start at approximately 0.5~m apart.  
    The initial desired relative position is $|d_{ij}|=0.4$~m, and at 60~s the desired relative position is switched to $|d_{ij}|=0.5$~m.}
\end{figure*}

\Cref{fig:cl_multiset_sim_2} shows the simulation results, which qualitatively matches the experimental results in \Cref{fig:cl_multiset_2}. 
In the experiment, the current amplitude saturation \Cref{eq:hat_I_ij} is active from $5.1$~s to $6.3$~s.
During this time interval, the unsaturated current amplitudes exceed the limit $\bar{I}=2.35$~A.
The peak unsaturated amplitude is $2.77$~A, which occurs at $5.3$~s and corresponds to a desired force of $0.0078$~N.
Since the amplitude is saturated, the time-averaged force is only $0.0066$~N, which is 15\% less than the desired value. 
However, the the formation remains stable with only a small increase in settling time.
Specifically, the current saturation causes less that 3\% increase in settling time (determined by comparing the saturated versus unsaturated simulations).
Collectively, this demonstrates that the control approach has some robustness to the current amplitude limits imposed by the hardware.
\exampletriangle
\end{experiment}


\subsection{Three-Satellite Closed-Loop Formation Experiments}

Next, we present 3 closed-loop experiments with $n=3$ satellites. 
The middle, left, and right satellites are numbered~1, 2, and 3. 
The neighbor sets are $\mathcal{N}_{1}=\{2,3 \}$, $\mathcal{N}_{2}=\{1 \}$, and $\mathcal{N}_{3}=\{1 \}$. 
The interaction frequencies are selected as the 2 smallest common multiples of $2\pi/T$, where $T=0.1$~s. 
Specifically, $\omega_{12}=\omega_{21}= 20 \pi$~rad/s and $\omega_{13}=\omega_{31}= 40 \pi$~rad/s. 
The smallest common multiples are used because lower frequency results in smaller impedance and larger maximum allowable current.
Recall that the amplifiers can reproduce the above frequencies without generating common harmonics that could compromise the accuracy of the time-averaged force calculation (see \Cref{fig:freq_specbench_tests}).

We implement the currents $(I_{ij,k}, I_{ji,k})$ given by  \eqref{eq:f*_ij_1D}--\eqref{eq:hat_I_ij} with $\epsilon_0=0.015$ m, $\epsilon_1=0.021$ m, $\alpha_{ij}=0.0158$, $\beta=7.38$, $\rho_{ij}=0.00136$, $\gamma_{12} = \gamma_{13} = 0.8$ and $ \gamma_{21}=\gamma_{31}=1.25$, and $\bar{I}=2.35$ A.
Note that we select $\alpha_{ij}$, $\beta$, and $\rho_{ij}$ such that $\alpha_{ij}m/(c_0 N^2 A^2)=812$, $\alpha_{ij} \beta m/(c_0 N^2 A^2)=6000$, and $\rho_{ij} m/(c_0 N^2 A^2) =70$. 
The gains are selected based on those used in the 2-satellite experiments and in 3-satellite simulations.
In particular, initial values of $\alpha_{ij}$ and $\beta$ were the same as in the 2-satellite case. 
The integral gain $\rho_{ij}$ was increased from zero until the maximum SSE in simulation was less than 0.005~m. 
However, the nonzero integral gain increases settling time; thus, $\beta$ was increased by 7\% to reduce the simulation settling time to no more than 50~s. 
Similar to the 2-satellite case, there is a tradeoff between response characteristics (e.g., settling time, overshoot) and hardware capabilities (e.g., current limits, sensor noise).
Increasing $\beta$ beyond the value used here degrades performance because it amplifies noise in the velocity-feedback term.

\begin{figure*}[!t]
    \begin{subfigure}{0.49\textwidth}
        \centering
        \includegraphics[width=1\textwidth,clip=true,trim= 0.0in 0.5in 0in 0.5in]{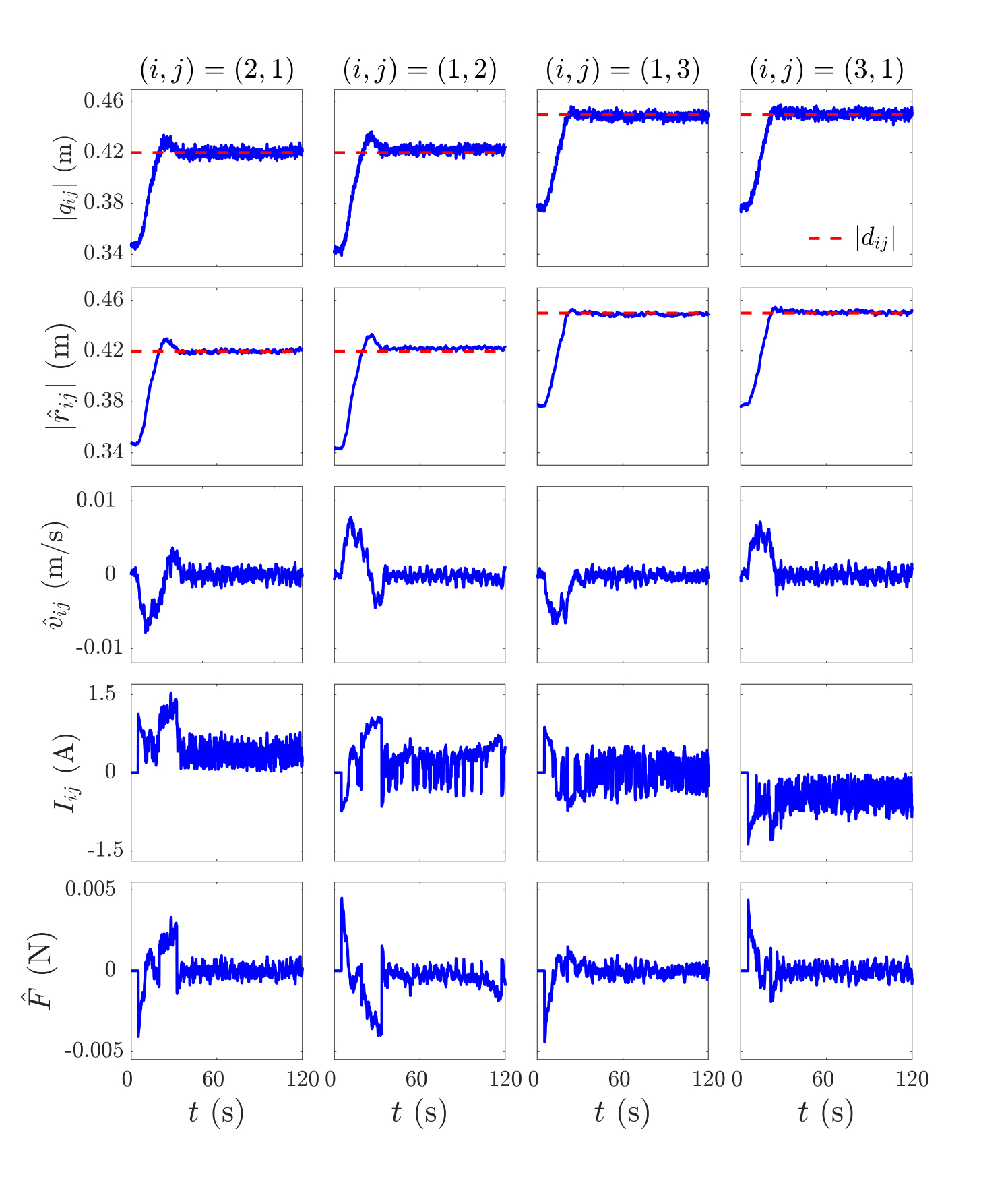}
        \caption{Experiment}
        \label{fig:3_sat_repulsion_42_45_2}
    \end{subfigure}
    \hfill
    \begin{subfigure}{0.49\textwidth}
        \centering
        \includegraphics[width=1\textwidth,clip=true,trim= 0.0in 0.5in 0in 0.5in]{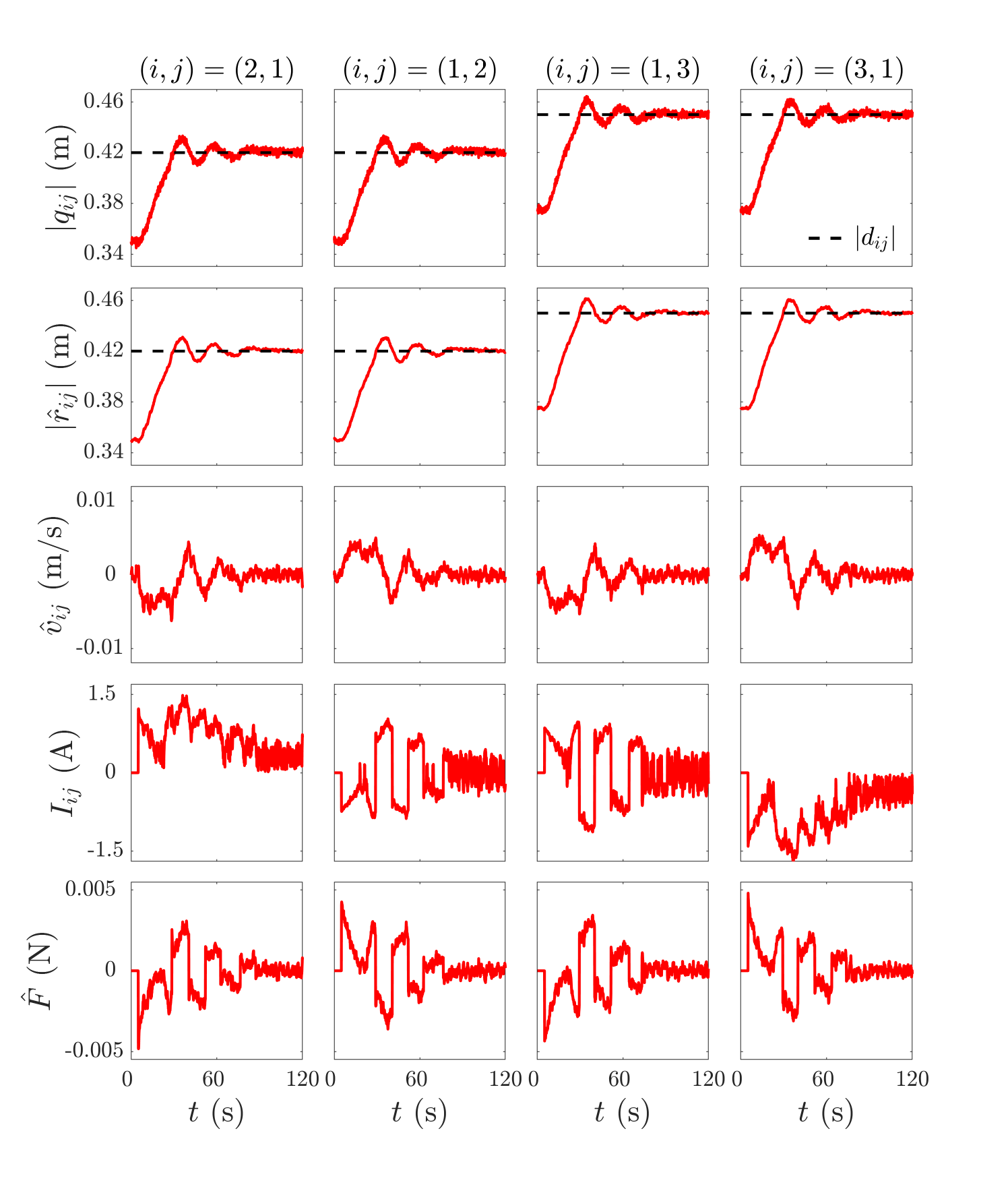}
        \caption{Simulation}
        \label{fig:3_sat_sim_repulsion_42_45_2}
    \end{subfigure}
    \caption{Closed-loop repulsion between satellite~1 and~2, and repulsion between satellite~1 and~3. 
    Initially, satellites~1 and~2 are approximately 0.346~m apart, and satellites~1 and~3 are approximately 0.377~m apart. 
    The desired relative positions are $d_{12}=0.42$~m and $d_{13}=-0.45$~m.}
\end{figure*}

\begin{experiment}\rm \label{exp:3_sat_repulsion} 
This experiment demonstrates closed-loop repulsion between satellites~1 and~2, and closed-loop repulsion between satellites~1 and~3. 
Satellites~1 and~2 start at a relative distance of approximately 0.35~m, and satellites~1 and~3 start at a relative distance of approximately 0.38~m. 
The objective is to reach the desired relative positions $d_{12}=0.42$~m and $d_{13}=-0.45$~m. 
\Cref{fig:3_sat_repulsion_42_45_2} shows the experimental results, where the column on the left shows relative measurements, estimates, and time-averaged approximate force between satellite~2 (left satellite) and satellite~1 (middle satellite), as measured and computed on satellite~2; and the second column from the left shows the data as measured and computed on satellite~1. 
The third column from the left shows relative measurements, estimates, and time-averaged approximate force between satellite~1 and satellite~3 (right satellite), as measured and computed on satellite~1; and the column on the right shows the data as measured and computed on satellite~3. 
\Cref{fig:3_sat_repulsion_42_45_2} shows that $\hat{r}_{12}$ and $\hat{r}_{13}$ converges to $d_{12}$ and $d_{13}$ by $30$~s. 
For satellites~1 and~2, the settling time is 25.4~s, and the mean SSE $| \hat{r}_{12}-d_{12} |$ over the last 30~s is $6.3 \times 10^{-4}$~m. 
For satellites~1 and~3, the settling time is 22.3~s, and the mean SSE $| \hat{r}_{13}-d_{13} |$ over the last 30~s is $6.8 \times 10^{-4}$~m.
The variance of the unfiltered SSE $ q_{ij}-d_{ij} $ is $4.3\times 10^{-6}$~m$^2$, where $2\times 10^{-6}$ is attributed to sensor noise variance.
The remaining variance is potentially due in part to unmodeled external disturbances as well as unmodeled dynamics, which can influence SSE through feedback as discussed in \Cref{exp:2_sat_repulsion}.

Next, we present simulations of the closed-loop \Cref{accel_i,u_i} where $\mathbf{p}_{ij,k}=p_{ij,k} \mathbf{i}$. 
The effect of friction between the satellites and the linear air tracks becomes noticeable with the addition of the third satellite. 
This effect is modeled by including linear damping $b_i v_i \mathbf{i}$ in \Cref{accel_i}, where selecting $b_i$ involves a tradeoff between reducing response oscillations and increasing peak time (i.e., time of largest overshoot peak). 
Larger $b_i$ tends to decrease oscillations but increases peak time.
The damping coefficient $b_i = 0.08$~N-s/m is selected to achieve a similar response to the experiments.
In particular, $b_i$ is selected such that the overshoot in simulations is within 40\% of the experiment for all 3 experiments in this section. 
However, this comes with the tradeoff of larger peak time. 
Across all experiments, the simulation peak time is within 35\% of the experiment.
The simulated relative position includes zero-mean Gaussian-white sensor noise with variance $V=2\times 10^{-6}$~m$^2$, which is the sensor noise variance estimated from experimental data.  
\Cref{fig:3_sat_sim_repulsion_42_45_2} shows the simulation results, which qualitatively agree with the experimental results. 
However, we observe an oscillatory transient response in \Cref{fig:3_sat_sim_repulsion_42_45_2}, which is not observed in \Cref{fig:3_sat_repulsion_42_45_2}. 
This can partly be explained by the imperfect friction model, unmodeled external disturbances (e.g., lateral air pressure from the air track), and unmodeled near-field electromagnetic effects.
\exampletriangle
\end{experiment}

\begin{figure*}[!t]
    \begin{subfigure}{0.49\textwidth}
        \centering
        \includegraphics[width=1\textwidth,clip=true,trim= 0.0in 0.5in 0in 0.5in]{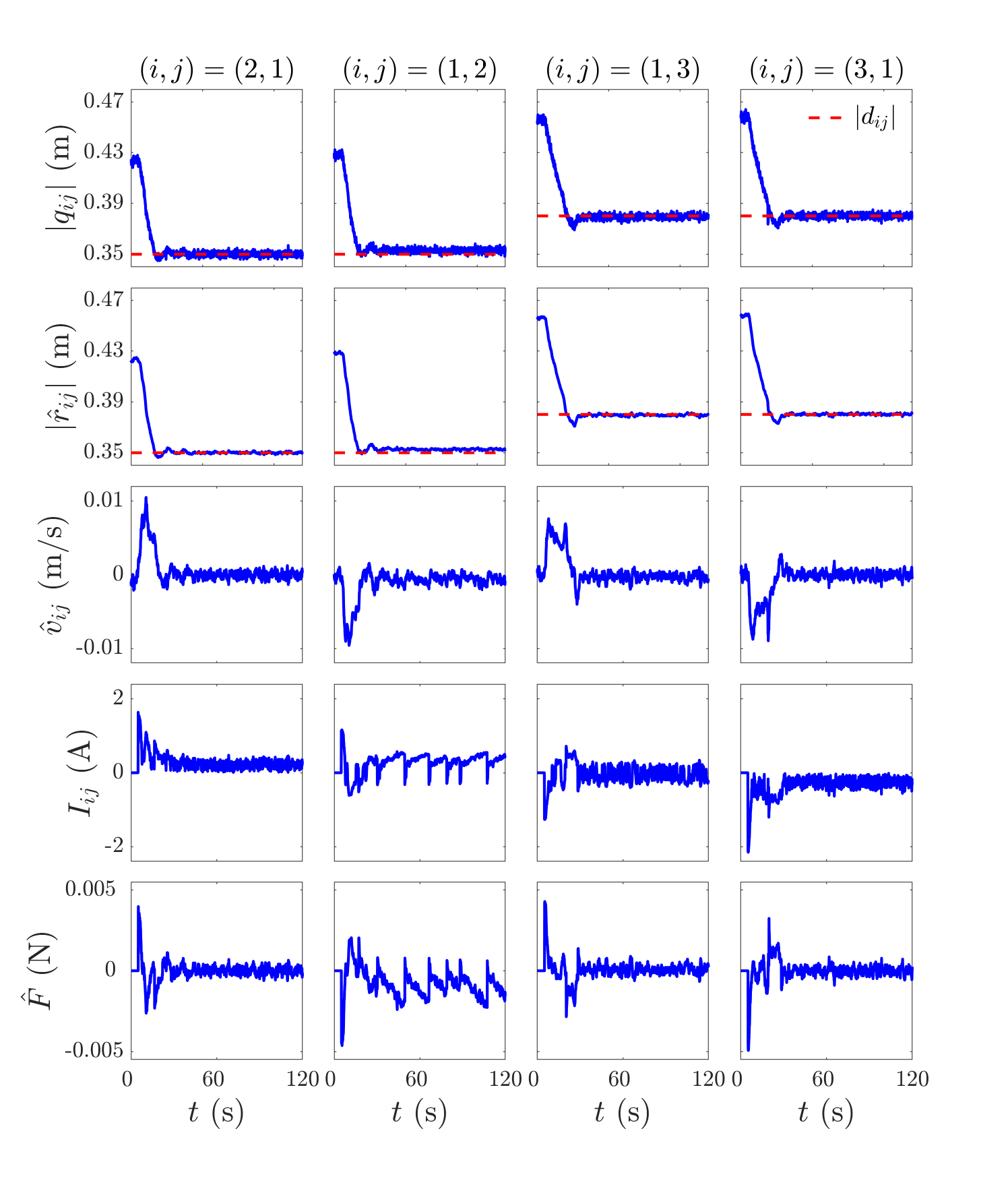}
        \caption{Experiment}
        \label{fig:3_satAttraction_35_38}
    \end{subfigure}
    \hfill
    \begin{subfigure}{0.49\textwidth}
        \centering
        \includegraphics[width=1\textwidth,clip=true,trim= 0.0in 0.5in 0in 0.5in]{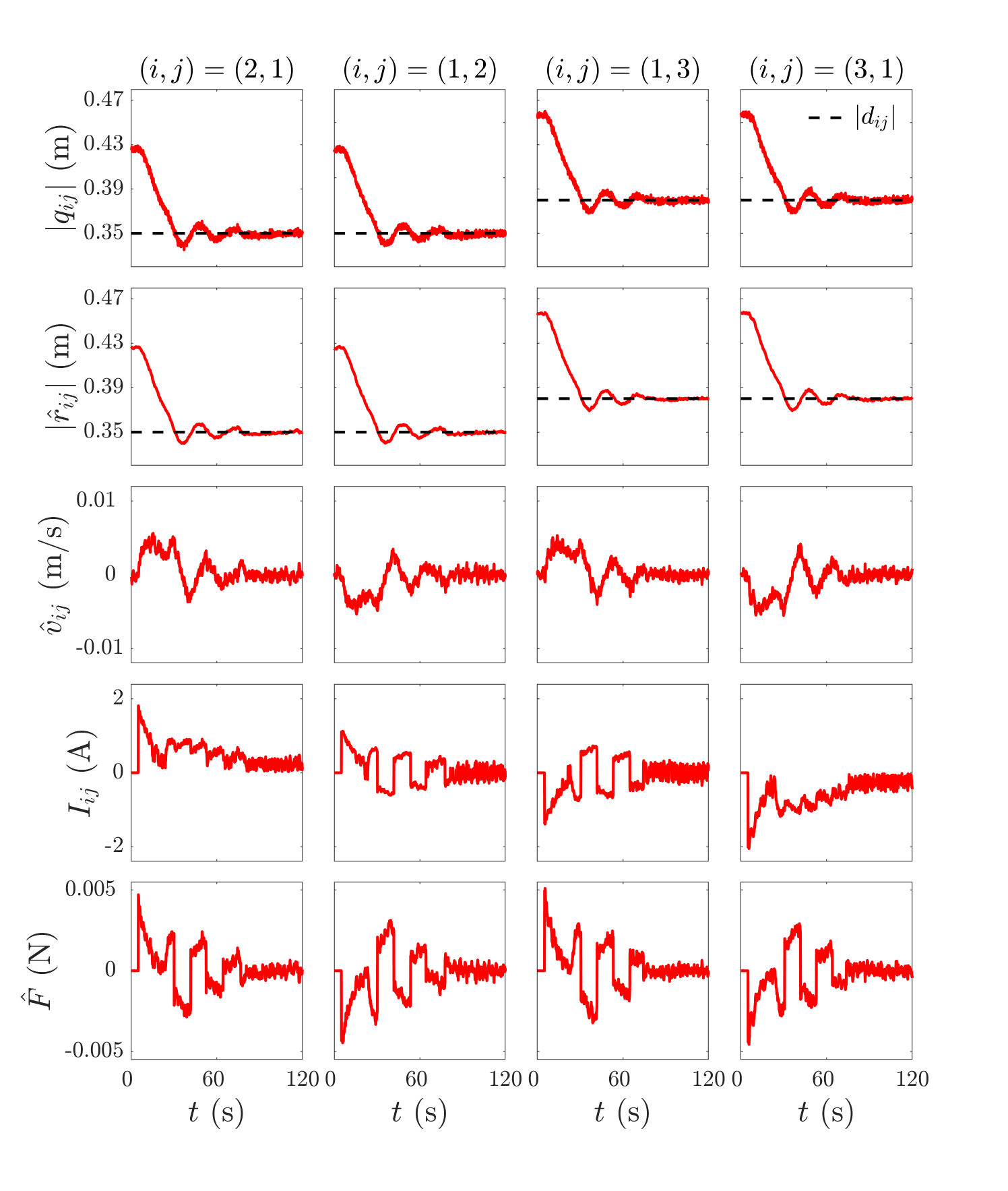}
        \caption{Simulation}
        \label{fig:3_sat_sim_Attraction_35_38}
    \end{subfigure}
    \caption{Closed-loop attraction between satellite~1 and~2, and attraction between satellite~1 and~3. 
    Initially, satellites~1 and~2 are approximately 0.425~m apart, and satellites~1 and~3 are approximately 0.46~m apart. 
    The desired relative positions are $d_{12}=0.35$~m and $d_{13}=-0.38$~m.}
\end{figure*}

\begin{experiment}\rm \label{exp:3_sat_attraction} 
This experiment demonstrates closed-loop attraction between satellite~1 and satellite~2, and closed-loop attraction between satellite~1 and satellite~3. 
Satellites~1 and~2 start at a relative distance of approximately 0.425~m, and satellites~1 and~3 start at a relative distance of approximately 0.46~m. 
The objective is to reach the desired relative positions $d_{12}=0.35$~m and $d_{13}=-0.38$~m. 
\Cref{fig:3_satAttraction_35_38} shows that $\hat{r}_{12}$ and $\hat{r}_{13}$ converges to the desired relative positions $d_{12}$ and $d_{13}$ by $25$~s.
For satellites~1 and~2, the settling time is 17.2~s, and the mean SSE $|\hat{r}_{12}-d_{12}|$ over the last 30~s is $4.7 \times 10^{-4}$~m. 
For satellites~1 and~3, the settling time is 20.9~s, and the mean SSE $|\hat{r}_{13}-d_{13}|$ over the last 30~s is $5.3 \times 10^{-4}$~m.
The variance of the unfiltered SSE $q_{ij}-d_{ij}$ is $3.4\times 10^{-6}$~m$^2$, which is comparable to \Cref{exp:3_sat_repulsion}.
\Cref{fig:3_sat_sim_Attraction_35_38} shows the simulation results, which qualitatively agree with the experimental results. 
Similar to the comparison in \Cref{exp:3_sat_repulsion}, the simulation has oscillatory overshoot whereas the overshoot in the experiment decays more quickly. 
\exampletriangle
\end{experiment}

\begin{figure*}[h!]
    \begin{subfigure}{0.49\textwidth}
        \centering
        \includegraphics[width=1\textwidth,clip=true,trim= 0.0in 0.5in 0in 0.5in]{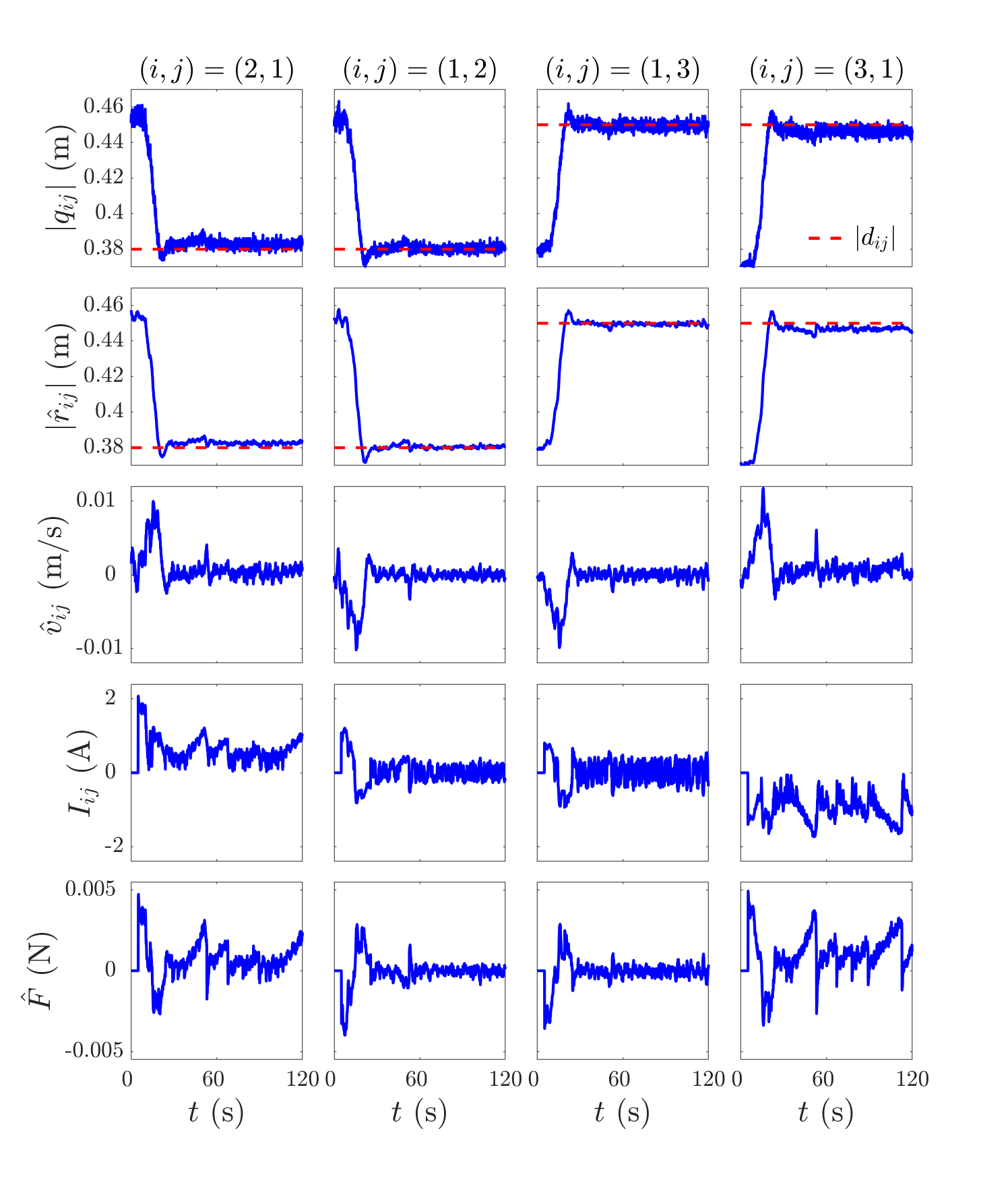}
        \caption{Experiment}
        \label{fig:3_satAttraction_Repulsion_38_45}
    \end{subfigure}
    \hfill
    \begin{subfigure}{0.49\textwidth}
        \centering
        \includegraphics[width=1\textwidth,clip=true,trim= 0.0in 0.5in 0in 0.5in]{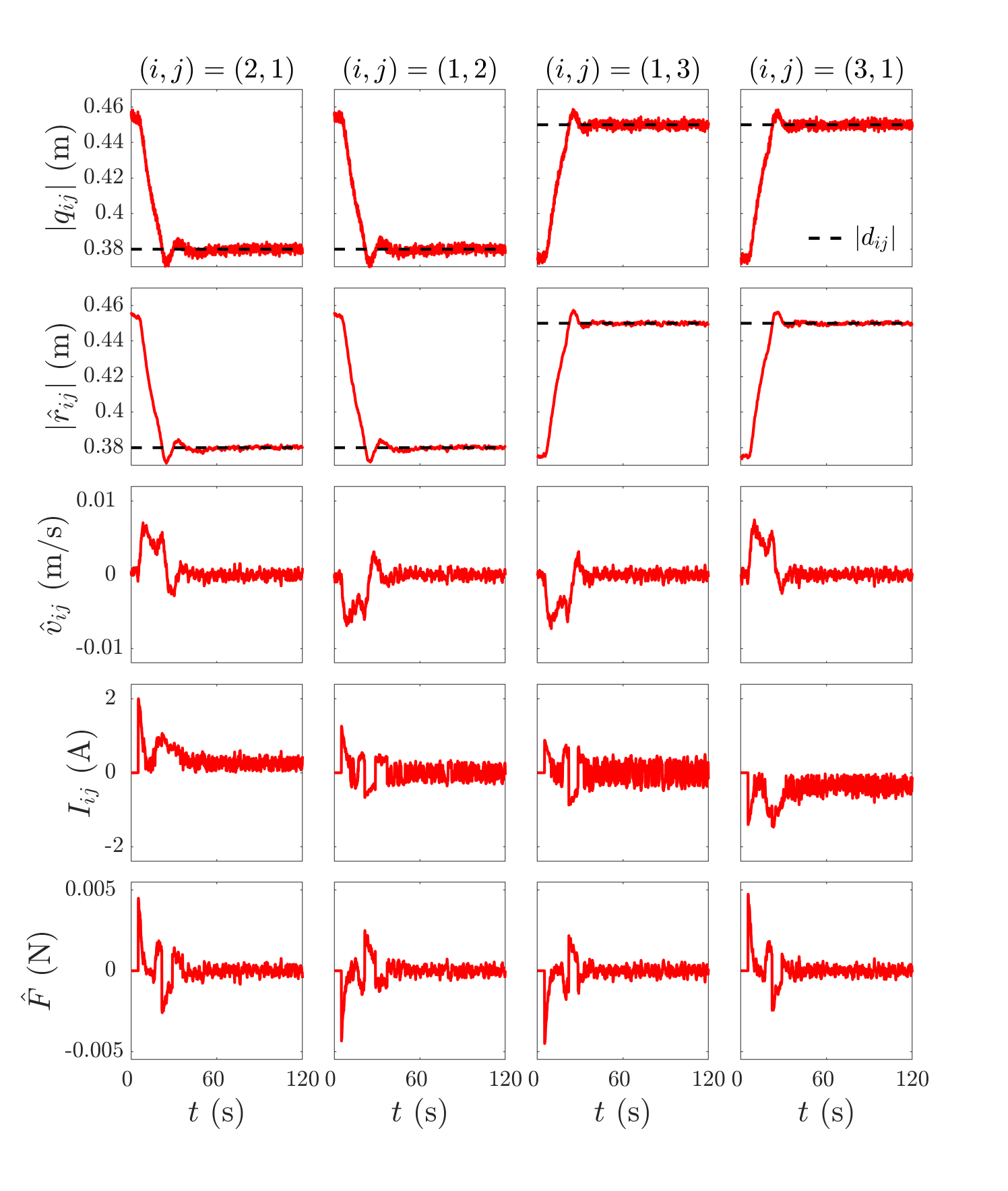}
        \caption{Simulation}
         \label{fig:3_sat_sim_Attraction_Repulsion_38_45}
    \end{subfigure}
    \caption{Closed-loop attraction between satellite~1 and~2, and repulsion between satellite~1 and~3. 
    Initially, satellites~1 and~2 are approximately 0.46~m apart, and satellites~1 and~3 are approximately 0.38~m apart. 
    The desired relative positions are $d_{12}=0.42$~m and $d_{13}=-0.45$~m.}
\end{figure*}


\begin{experiment}\rm \label{exp:3_sat_attraction_repulsion} 
This experiment demonstrates closed-loop attraction between satellite~1 and satellite~2, and closed-loop repulsion between satellite~1 and satellite~3. 
Satellites~1 and~2 start at a relative distance of approximately 0.46~m, and satellites~1 and~3 start at a relative distance of approximately 0.38~m.
The objective is to reach the desired relative positions $d_{12}=0.42$~m and $d_{13}=-0.45$~m. 
\Cref{fig:3_satAttraction_Repulsion_38_45} shows that $\hat{r}_{12}$ and $\hat{r}_{13}$ converges to the desired relative position $d_{12}$ and $d_{13}$ by $25$~s. 
For satellites 1 and 2, the settling time is 22.4~s, and the mean SSE $| \hat{r}_{12}-d_{12} |$ over the last 30~s is $5.6 \times 10^{-4}$~m. 
For satellites 1 and 3, the settling time is 22.6~s, and the mean SSE $|\hat{r}_{13}-d_{13}|$ over the last 30~s is $6.3 \times 10^{-4}$~m.
The variance of the unfiltered SSE $q_{ij}-d_{ij}$ is $3.4\times 10^{-6}$~m$^2$, which is comparable to \Cref{exp:3_sat_repulsion} and \Cref{exp:3_sat_attraction}.
\Cref{fig:3_sat_sim_Attraction_Repulsion_38_45} shows the simulation of the experiment, which qualitatively matches the experimental results in \Cref{fig:3_satAttraction_Repulsion_38_45}. 
\exampletriangle
\end{experiment}

\Cref{table:exp_results_summ} summarizes experimental metrics, including settling time, maximum SSE $|\hat{r}_{ij}-d_{ij}|$ over the last 30~s, mean SSE, and the variance of the SSE.

\begin{table}[h!]
\caption{Settling Time, Maximum SSE, Mean SSE, and Variance of SSE}
\begin{center}
\resizebox{\columnwidth}{!}{%
\begin{tabular}{c|c|c|c|c}
 & &Maximum & Mean & Variance\\
Experiment&  $T_{\rms}$ (s) &SSE (m)& SSE (m) & of SSE (m$^2$)\\
\hline
3   &19 & $2.7 \times 10^{-3}$ & $7.7 \times 10^{-4}$  &9.3 $\times 10^{-7}$   \\ 
4   &20  &$1.7 \times 10^{-3}$ & $5.3 \times 10^{-4}$   &4.3 $\times 10^{-7}$  \\  \hline
6   &24  &$2.7 \times 10^{-3}$   & $6.5 \times 10^{-4}$   &6.7 $\times 10^{-7}$  \\ 
7   &20  &$2.3 \times 10^{-3}$  &$4.9 \times 10^{-4}$    &3.8 $\times 10^{-7}$  \\ 
8   &23  &$2.6 \times 10^{-3}$  & $5.9 \times 10^{-4}$   &5.7 $\times 10^{-7}$   \\ 
\end{tabular}%
}
\end{center}
\label{table:exp_results_summ}
\end{table}

\begin{figure*}[t!]
\begin{subfigure}{0.49\textwidth}
    \centering
    \includegraphics[width=1\textwidth,clip=true,trim= 0.3in 0.1in 0.5in 0.2in]{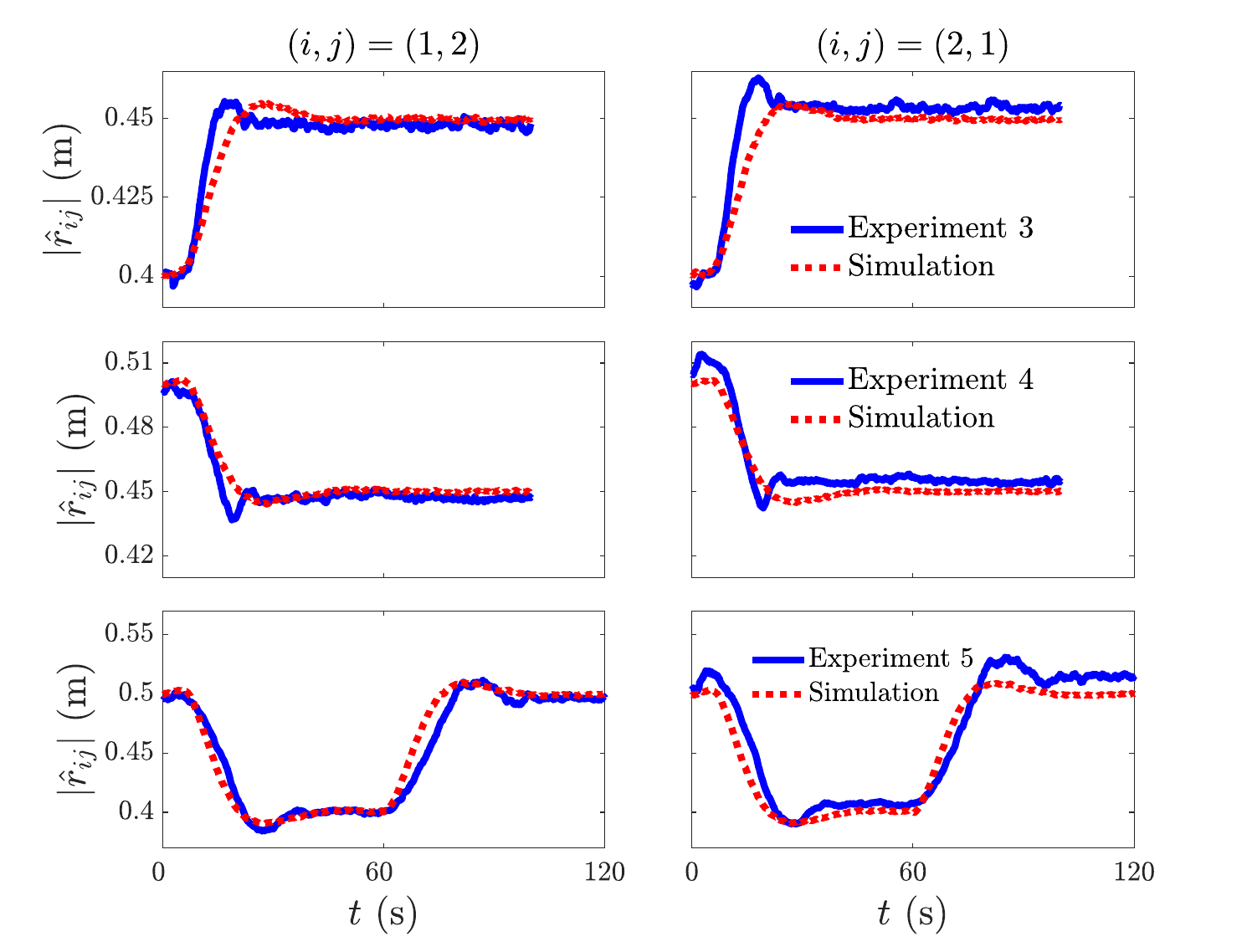}
    \caption{Filtered relative position $\hat{r}_{ij}$.}    \label{fig:2_sat_relative_position_combined}
    \end{subfigure}
    \hfill
\begin{subfigure}{0.5\textwidth}
    \centering
    \includegraphics[width=1\textwidth,clip=true,trim= 0.15in 0.1in 0.5in 0.2in]{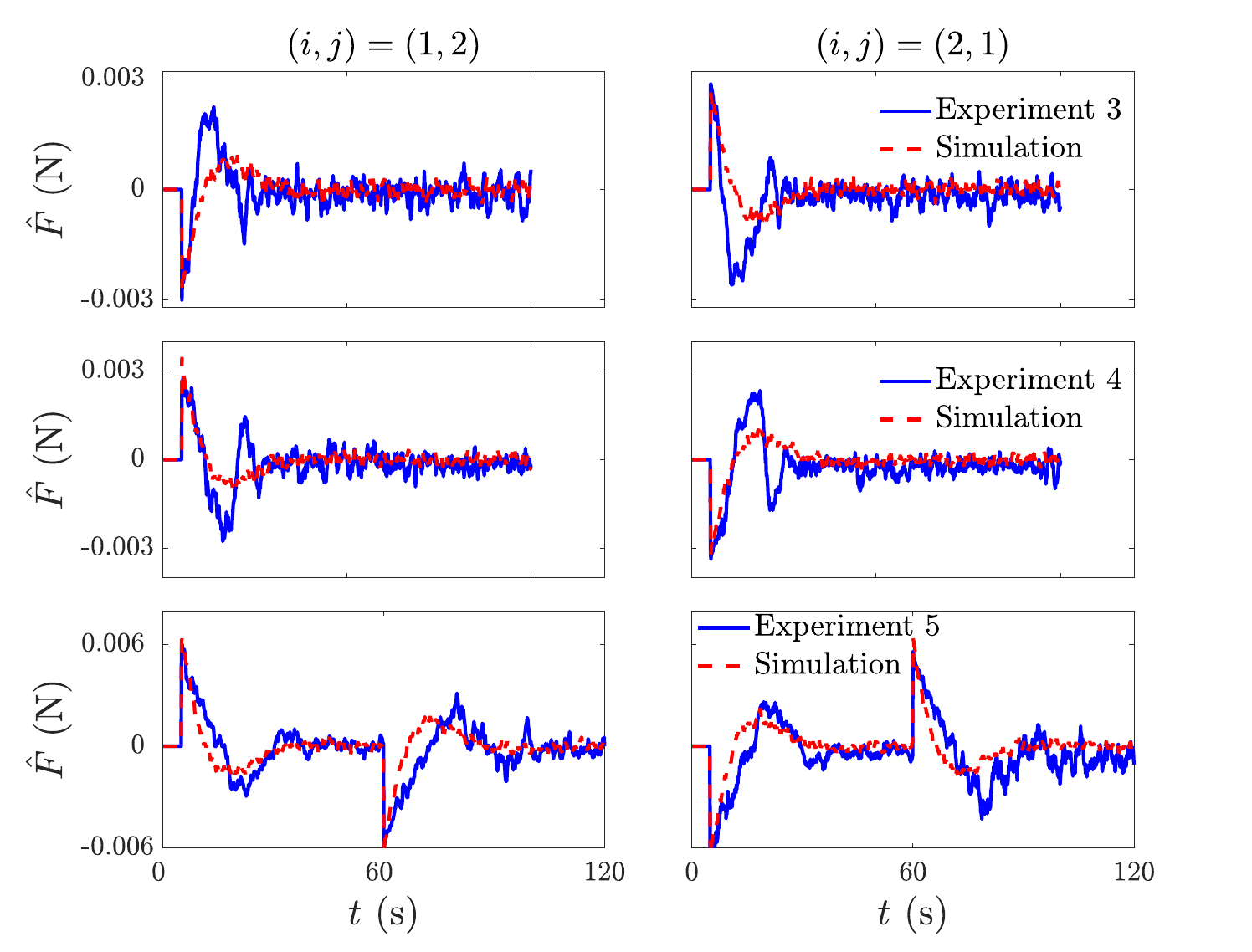}
    \caption{Time-averaged intersatellite force $\hat{F}$.}
    \label{fig:2_sat_force_combined}
    \end{subfigure}
    \caption{Comparison of 2-satellite experiments to simulations.}
        \label{fig:2_sat_comp}
\end{figure*}
\begin{figure*}[t!]
    \begin{subfigure}{0.49\textwidth}
    \centering
    \includegraphics[width=1\textwidth,clip=true,trim= 0.3in 0.1in 0.5in 0in]{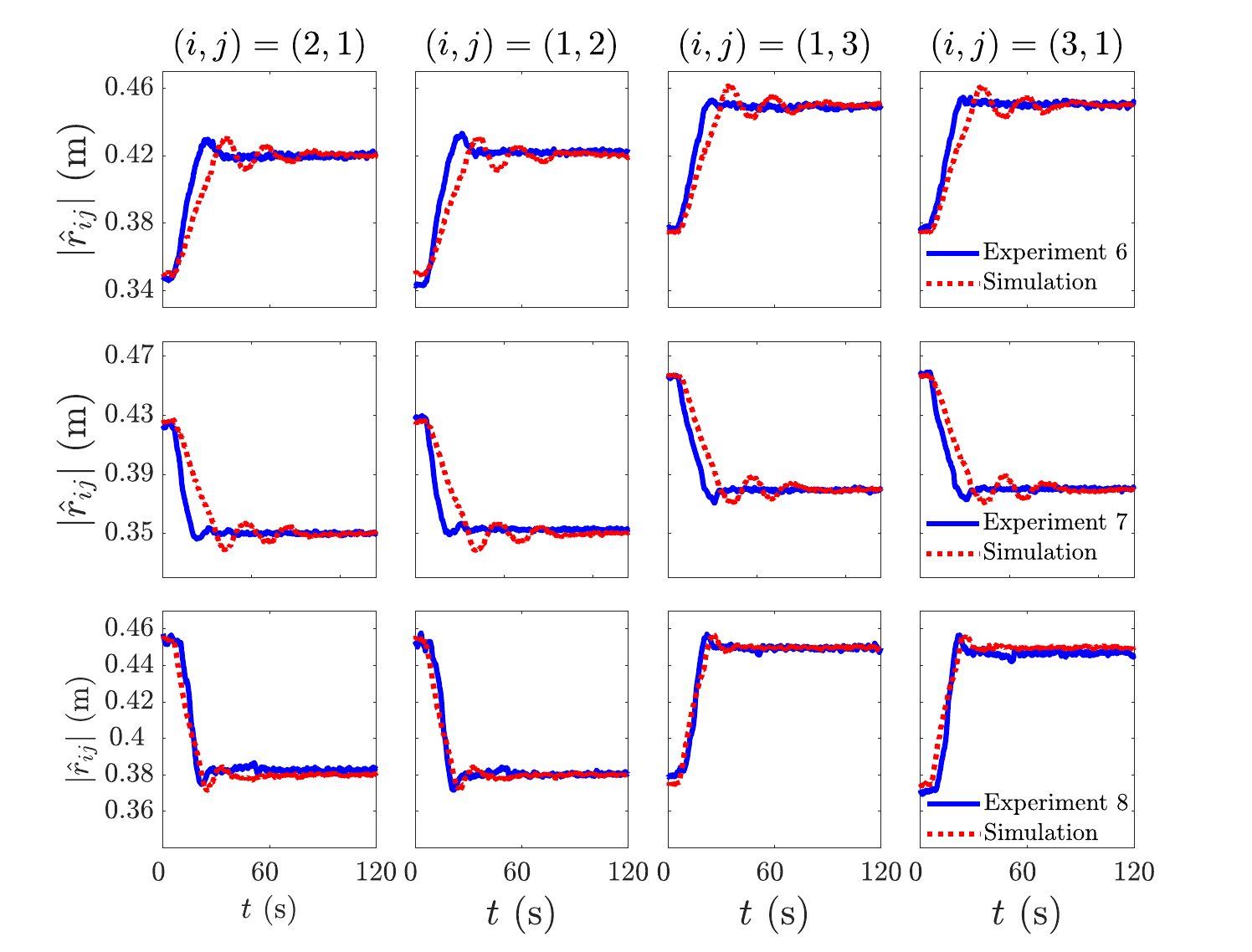}
    \caption{Filtered relative positions $\hat{r}_{ij}$.}
    \label{fig:3_sat_relative_position_combined}
    \end{subfigure}
    \hfill
    \begin{subfigure}{0.5\textwidth}
    \centering
    \includegraphics[width=1\textwidth,clip=true,trim= 0.1in 0.1in 0.5in 0.2in]{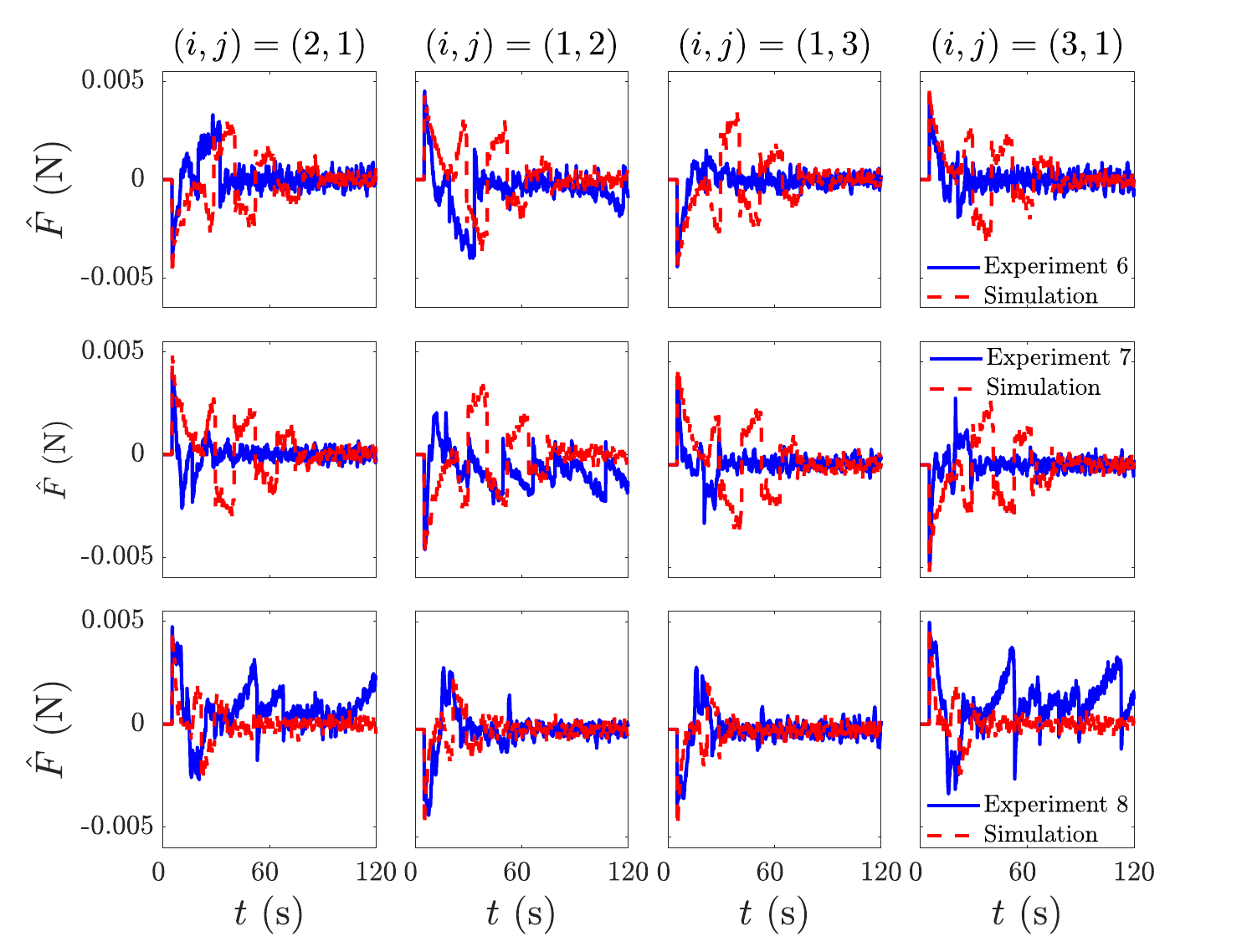}
    \caption{Time-averaged intersatellite force $\hat{F}$.}
    \label{fig:3_sat_force_combined}
    \end{subfigure}
    \caption{Comparison of 3-satellite experiments to simulations.}
    \label{fig:3_sat_comp}
\end{figure*}

\subsection{Experiment-to-Simulation Comparison}

\Cref{fig:2_sat_comp,fig:3_sat_comp} provide a comparison of $\hat r_{ij}$ and $\hat F_{ij}$ for the 2- and 3-satellite cases.
These figures demonstrate good qualitative agreement between the experiments and simulations particularly for relative position. 
The intersatellite force $\hat F_{ij}$ has larger peak-to-peak steady-state variations in the experiments (particularly 3 satellites), which can be explained by the feedback compensating for unmodeled external disturbances (e.g., lateral air pressure from the air track) in the experiment that are not included in the simulation. 
For \Cref{exp:3_sat_repulsion} and \Cref{exp:3_sat_attraction}, the simulation response has more oscillations and large peak time than observed in the experiments. 
The damping parameter $b_i$ can be increased to reduce the oscillations in the experiment; however, this makes peaks time even larger. 
This can partly be explained by an imperfect friction model, unmodeled external disturbances, and unmodeled near-field electromagnetic effects.

\begin{figure}[hbt!]
    \centering
    \includegraphics[width=0.48\textwidth,clip=true,trim= 0.2in 1in 0.8in 0.65in]{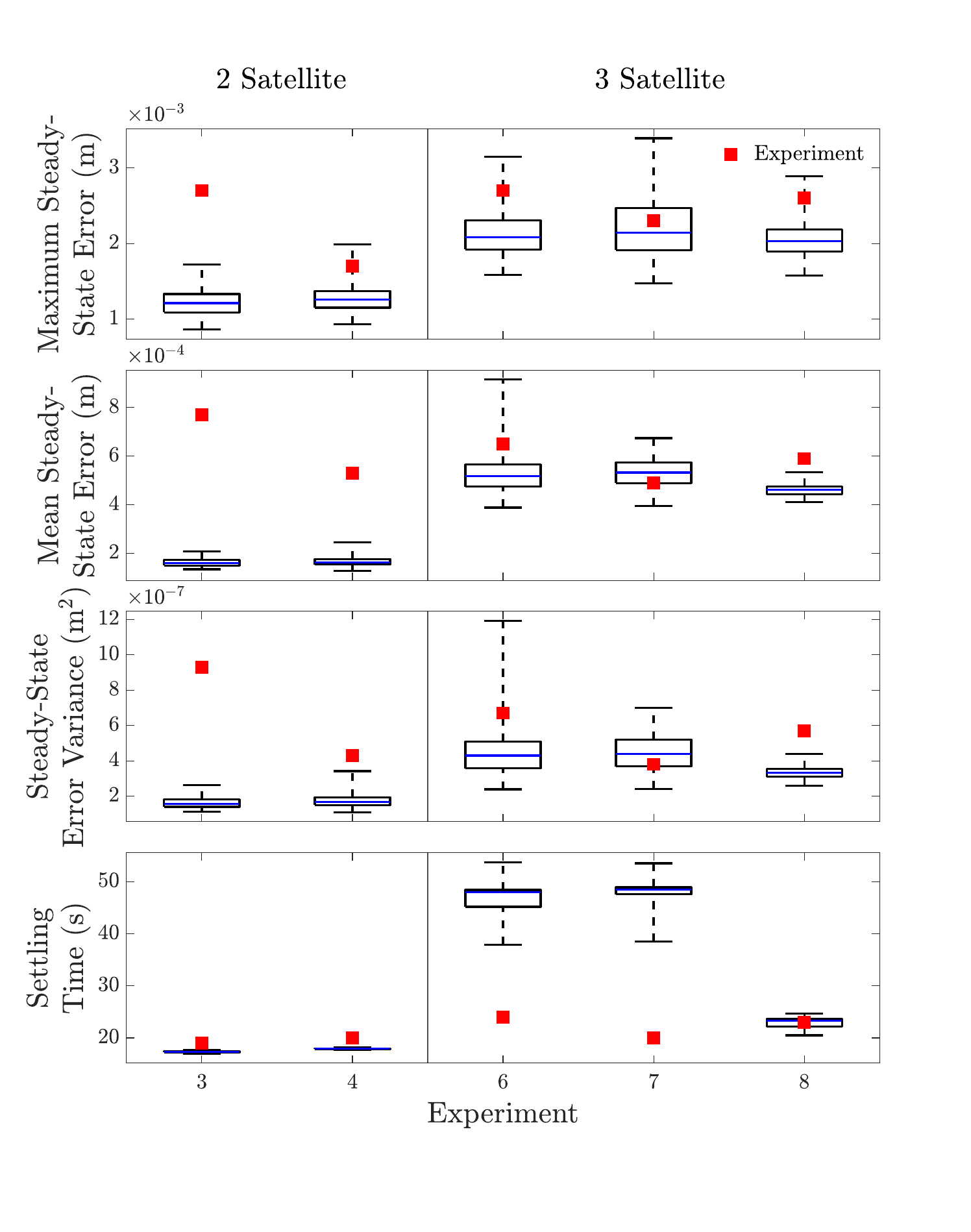}
        \caption{Maximum SSE, mean SSE,  variance of SSE, and settling time for experiments and 100 simulations. 
        Experimental values are shown by the red box. 
        The box-and-whisker plot show the median (blue line), middle two  quartiles (box), and full range (whisker) for the 100 simulations.
        }
        \label{fig:metrics_experiment_simulations}    
\end{figure}

Next, we compare the maximum SSE, mean SSE, variance of the SSE, and settling time in the experiments to those observed in $100$ simulations with different noise realizations.   
\Cref{fig:metrics_experiment_simulations} provides a box-and-whisker plot for the metrics on all 100 simulations as well as the metrics from the experiments. 
The SSE metrics are the same order of magnitude across all experiments and simulations. 
For the 3-satellite experiments, the SSE metrics are inside the simulation range or close.
Notably, the maximum SSE is inside the simulation range for all experiments except \Cref{exp:2_sat_repulsion}.

The settling time for \Cref{exp:3_sat_attraction} and \Cref{exp:3_sat_repulsion} differ noticeably from simulation  (i.e., 50\% less than simulation mean). 
One potential explanation is the use of a linear friction model in simulation. 
The friction in the experiment is likely to have nonlinear effects; for example, the friction force may decay sublinearly with velocity or have Coulomb behavior. 
Sublinear or Coulomb friction could be used to model a friction forces that are larger than linear friction at low velocity and smaller than linear friction at high velocity. 
This more complex friction model may be able to capture the fast peak- and rise-time behavior of the experiments while damping out steady-state oscillations to have a low settling time, similar to those observed in the experiments.

%

\section{Concluding Remarks}

This article presented a 3-satellite experimental demonstration of decentralized EMFF using AMFF.
To the authors' knowledge, this is the first demonstration of AMFF with at least 3 satellites in open or closed loop. 
The experimental results not only demonstrate the feasibility of EMFF with AMFF but also demonstrates that the frequency-multiplexed amplitude-modulated sinusoidal approach can achieve formation with multiple satellites. 
Specifically, the closed-loop experiments demonstrate decentralized EMFF with AMFF for 3 satellites where the maximum steady-state formation error is less than $\pm 0.01$~m, the mean steady-state formation error is less than $\pm 0.001$~m, and the settling time is less than $30$~s.
These experiments validate the decoupling of intersatellite forces through frequency-multiplexed AMFF.
The settling time and steady-state formation error achieved in these experiments demonstrate that this approach is most likely capable of meeting requirements for satellite constellation reconfiguration and maintenance, thereby addressing the need for propellant-free formation flying.

\subsection{Strengths of Frequency-Multiplexed AMFF}

This article demonstrates several strengths of frequency-multiplexed AMFF, including: (1) the decoupling of intersatellite forces for $n > 2$; (2) decentralized implementation; (3) scalable to a large number of satellites; and (4) robustness to hardware constraints. 
The core concept of AMFF is the time-averaged force decoupling that results from the use of frequency-multiplexed sinusoids, where the amplitude can be modulated (i.e., controlled) to achieve desired intersatellite forces.  
This feature unlocks decentralized and scalable implementation.

Frequency-multiplexed AMFF is well suited to decentralization.
In contrast, we are unaware of an effective decentralized implementation of DC-driven EMFF. 
If the established DC-driven approach (e.g., \cite{Schweighart2010}) is implemented in a decentralized architecture (i.e., each satellite determines its DC current from a nonconvex constrained optimization involving only its neighbors), then the formation is generally unstable. 
Specifically, this decentralized DC-driven implementation results in an unstable (diverging) formations applied to the simple 3-satellite single-degree-of-freedom scenario in this article.

Next, we note that AMFF easily scales to large formations. 
In fact, scaling is related to the number of neighbors rather than the total number of satellites because AMFF is decentralized. 
Notably, the number of unique frequencies and the complexity of the algorithm in \Cref{sec:Abbasi_Algorithm} scales linearly with the number of satellites in the neighbor set. 
In contrast, DC-driven EMFF (centralized implementation) typically requires solving a constrained nonconvex optimization that suffers from the curse of dimensionality and does not scale well to large formations. 
On the one hand, a centralized DC-driven implementation yields formation results that are similar to the decentralized AMFF simulations in this paper (e.g., \Cref{fig:3_sat_sim_repulsion_42_45_2,fig:3_sat_sim_Attraction_35_38,fig:3_sat_sim_Attraction_Repulsion_38_45}). 
However, the optimization to determine coil currents is not guaranteed to converge, particularly as the number of satellites grows.

Lastly, experiments in this paper demonstrate that AMFF provides robustness to hardware limitations (e.g., power limits, sensor noise). 
Nevertheless, future experimental studies could adopt the AMFF control methods in \cite{Kamat2025b} to explicitly account for constraints (e.g., power/current limits, thermal management of the coils) using a composite control barrier function.

\subsection{Limitations and Opportunities for Future Work}

The experiments in this article use low-friction ground-based air tracks. 
Future studies could focus on considerations that are relevant for in-space testing and deployment, including the impact of space environment (e.g., microgravity, radiation, extreme temperature cycles) and the design of space-ready hardware (e.g., coils, sensing package).

Other important considerations for on-orbit deployment include accounting for gravitational forces (which is addressed in \cite{Kamat2025b}), disturbances such as atmospheric drag in low Earth orbit, and other space environment effects (e.g., radiation, extreme heat-cold cycles, Earth's magnetic field). 
For example, the performance of the Kalman filter may degrade in the presence of these non-Gaussian disturbances. 
In this case, other state estimation methods (e.g., particle filters) could be considered.

Although there was no significant time or frequency drift observed in the experiments, this could be a concern for long-duration space missions, necessitating periodic resynchronization of the satellites or adaptive frequency tuning. 
Incorporating resynchronization into the AMFF approach is an area for future work.

The approach in this paper uses the far-field model, for close-proximity operations near-field magnetic effects may introduce errors (up to 20\%), which could limit applicability for this close-proximity formations. 
These near-field effects could be considered in future work on EMFF with AMFF.

The method in this article applies to 3-degree-of-freedom position control. However, the experiments are restricted to a single degree of freedom. 
Future experiments could include ground-based 3-degree-of-freedom experiments (planar translation and one-degree-of-freedom attitude control). 
Such experiments would require multiple electromagnetic coils (e.g., 2 orthogonal coils for control in the plane) and one additional frequency per interacting pair if relative attitude control \cite{Abbasi2020Scitech} is considered. 
Similarly, 3-degree-of-freedom control requires at least 3 electromagnetic coils.
In this case, hardware design is tightly interlinked with power requirements, computational complexity of the control algorithms, and acceptable performance levels.

\section*{Conflict of Interest Statement}
There is no conflict of interest.

\section*{Acknowledgments}
This work is supported in part by the National Aeronautics and Space Administration (80NSSC17M0040), the National Science Foundation (1932105), and the Air Force Office of Scientific Research (FA9550-20-1-0028).

We would like to thank Floyd Taylor, Herb Mefford, and Jonathan T. Williams at the University of Kentucky for their invaluable advice on improving the physical testbed.

\bibliographystyle{elsarticle-num} 

 \bibliography{1D_EMFF_EXP}
\end{document}